\documentclass[twocolumn]{aastex62}

\def\simgt{\lower.5ex\hbox{$\; \buildrel > \over \sim \;$}}
\def\simlt{\lower.5ex\hbox{$\; \buildrel < \over \sim \;$}}

\newcommand\teff{T$_{\rm eff}$}

\newcommand{\msun}{\ensuremath{\, {M}_\odot}}

\newcommand{\ocen}{$\omega$\,Cen}

\def\he3{$^3$He}

\graphicspath{{./}{figures/}}

\received{XXXX}
\revised{XXX}
\accepted{XXXX}
\submitjournal{ApJ}

\shorttitle{UV HST observations in NGC\,6402}
\shortauthors{D'Antona et al.}

\begin{document}

\title{\bf HST observations of the globular cluster NGC\,6402 (M\,14) and its  peculiar multiple populations}
\correspondingauthor{Francesca D'Antona}
\email{francesca.dantona@inaf.it; franca.dantona@gmail.com}

\author{Francesca D'Antona}
\affiliation{INAF-- Osservatorio Astronomico di Roma, Via Frascati 33, 00078 Monteporzio Catone (Roma), Italy}
\author{Antonino P. Milone}
\affiliation{ Dipartimento di Fisica e Astronomia ``Galileo Galilei'', Univ. di Padova, Vicolo dell'Osservatorio 3, Padova, IT-35122}
\author{Christian I. Johnson}
\affiliation{Space Telescope Science Institute, 3700 San Martin Dr., Baltimore, MD 21218, USA}
\author{Marco Tailo} 
\affiliation{ Dipartimento di Fisica e Astronomia ``Galileo Galilei'', Univ. di Padova, Vicolo dell'Osservatorio 3, Padova, IT-35122}
\affiliation{ Dipartimento di Fisica e Astronomia ``Augusto Righi", Universit\`a degli Studi di Bologna, Via Gobetti 93/2, I-40129 Bologna, Italy}
\author{Enrico Vesperini}
\affiliation{Department of Astronomy, Indiana University, Bloomington, IN 47401, USA}
\author{Vittoria Caloi}
\affiliation{INAF -- IAPS Roma, Via Fosso del Cavaliere, Roma, Italy, IT-00133 }
\author{ Paolo Ventura} 
\affiliation{INAF-- Osservatorio Astronomico di Roma, Via Frascati 33, 00078 Monteporzio Catone (Roma), Italy}
\author{ Anna Fabiola Marino} 
\affiliation{INAF-Osservatorio Astrofisico di Arcetri, Largo E.Fermi, 5. 50125, Firenze, Italy}
\author{Flavia Dell'Agli}
\affiliation{INAF-- Osservatorio Astronomico di Roma, Via Frascati 33, 00078 Monteporzio Catone (Roma), Italy}


\begin{abstract}
We present {\it Hubble Space Telescope} ({\it HST}) photometric results for  NGC\,6402, a highly reddened very luminous Galactic globular cluster (GC). Recent spectroscopic observations of its red giant stars have shown a quite peculiar behavior in the chemistry of its multiple populations.
These results have prompted UV and optical  {\it HST} observations aimed at obtaining the cluster's ``Chromosome map" (ChM),  an efficient tool to classify GCs and characterize their multiple populations. 
We find that the discontinuity in the abundance distributions of O, Mg, Al and Na   inferred from spectroscopy is  more nuanced in the ChM, which is mostly sensitive to nitrogen.
Nevertheless, photometry in optical bands reveals a double main sequence, indicating a discontinuity in the helium content of the populations.  The population with the largest chemical anomalies (extreme) peaks at a helium mass fraction Y$\sim$0.31. This helium content is consistent with results from the analysis of  the distribution of horizontal-branch stars and the spectrophotometry of the red giants. The ChM and the color magnitude diagrams  are compared with those in NGC\,2808, a prototype GC with helium abundances up to Y$\gtrsim $0.35, and both
confirm that  NGC\,6402 does not host stellar populations with such extreme helium content. 
Further, the ChM reveals the presence of a group of stars with larger metallicity,  thus indicating that NGC\,6402 is a Type II cluster.  The modalities of formation of the multiple populations in NGC\,6402 are briefly surveyed, with main attention on the Asymptotic Giant Branch and Supermassive star models, and on possible clusters' merging. 
\end{abstract}

\keywords{stars: evolution; Globular Clusters}

\section{Introduction} 
\label{sec:intro}
In spite of being the tenth cluster in terms of luminosity (and mass) among the Galactic globular clusters (GC), NGC\,6402 (M\,14) has not been subject to close scrutiny like other massive GCs, being located close to the galactic plane and highly reddened (E(B-V)$\simeq$0.6). The cluster has a moderately high metallicity of [Fe/H]=$-1.13\pm0.05$ \citep{johnson2019}, which is similar to that of NGC\,2808, the cluster prototype for the study of multiple populations (MPs).  HST photometry from the \cite{piotto2002} snapshot survey showed that NGC\,6402 is an example of a ``second--parameter" cluster \citep[for a recent discussion, see][]{tailo2020}, as its  horizontal branch (HB) morphology is too blue for its metallicity. In fact, the HB includes only a few stars on the red side of the RR\,Lyr gap and extends to high \teff.  The most complete ground based photometry (B and V) down to the upper main sequence (MS) is provided by \cite{contreraspena2013}. The analysis and  a catalogue of 110 RR\,Lyr stars is given by \cite{contreraspena2018}, who also summarize the possible indications that this cluster has an extragalactic origin. \\

The light element abundances in Galactic (but also in the extragalactic) GCs display large variations,  with typical anti--correlations indicating the presence of two or more chemically distinct groups of stars  \citep[see e.g.][for a recent comprehensive summary]{gratton2019}. While the elemental abundances in some stars are similar to those of halo stars having the same metallicity (population 1G), the majority of stars  show abundance patterns typical of  gas processed at very high temperature (T$\sim$30--100\,MK, depending on the locus of processing) by proton capture reactions (population 2G\footnote{The nomenclature here follows that of Milone et al. papers. 1G actually means `first generation' to be considered as the generation of stars from which any population with signatures of hot proton capture nucleosynthesis necessarily follows (second generation or 2G), either within $\sim$10$^6$\,yr (supermassive star model) or within $\sim$10$^8$\,yr (AGB model). Where confusion with previous definitions arises we will clarify them.}).  
The abundance of helium in the standard halo and in the 1G of GCs is settled on the Big Bang composition (Y$\sim$0.25 in mass fraction).  However, it is larger in the p-processed gas of the enriched populations and reaches values up to Y$\sim$0.35 in the `extreme' groups present in a few massive clusters, such as  \ocen\ and NGC\,2808. In these clusters, the 2G itself is split into separate groups, highlighted photometrically by splits of the MS due to the different values of Y in each of them \citep{bedin2004, dantona2005, lee2005, piotto2007, tailo2016, bellini2017}, and, in NGC\,2808, by the grouping of different abundance anomalies revealed by high dispersion spectroscopy \citep{carretta2015}. 
A standard value of Y$\sim$0.35 is quoted in the literature for the blue MS of NGC\,2808  \citep[e.g.][]{dantona2005}, Y=0.37 is derived for the more complex case of $\omega$ Cen \citep[e.g.][]{tailo2016}. 
\\

In recent years, photometry in the UV bands, especially in the HST bands  of the UV  WFC3/UVIS filters F275W and F336W, complemented with the blue filter F438W, has proven to be a powerful way of achieving information on the multiple populations,  thanks to the sensitivity of these filters to the C, N, and O abundance variations. The results of the {\it HST} UV Legacy survey \citep{piotto2015} permitted classifying GCs in terms of their ``Chromosome Maps" \citep[ChMs,][]{milone2017chromo}, a pseudo-two color diagram where stars belonging to the different populations cluster in different loci, as shown by the comparison with the results of high dispersion spectroscopy \citep{marino2019chem}. 
Thus spectra and HST UV photometry are useful complementary tools to investigate multiple populations and attempt to reach a full understanding of their formation.\\

Both tools have shown that the most massive GCs exhibit the most complex chemical patterns.  C--N and Na--O anticorrelations are commonly present in all GCs, but the most massive clusters also show  Mg--Al, and sometimes even Mg--K anticorrelation \citep[e.g.,][]{cohenkirby2012,mucciarelli2015,carretta2021}.  
A small fraction  of Galactic  GCs  show  significant iron abundance variations, with very significant cluster-to-cluster differences in the percentage of stars with higher metallicity and in the degree of iron enhancement \citep[see, e.g.][and references therein]{johnson2015}. These ``iron--complex" clusters, discovered by spectroscopy and by the split RGB  in the hk narrow-band photometry \citep[e.g.][]{lee2015, lim2015b},  have been also identified by their remarkably complex ChM, as the higher metallicity stars are located at the red side of the standard ChM loci, and they were dubbed `Type II' clusters  \citep{milone2017chromo}.
Most Type II clusters have HB morphologies that extend to very high temperatures, which commonly identifies these stars as descendants of the high--helium population \citep[e.g.][]{dc2004, dc2008}. Finally, the more massive clusters contain a larger fraction, up to more than 80\%, of stars belonging to the anomalous 2G \citep[see Fig.\,22 in][]{milone2017chromo}. \\   

There is consensus that the gas forming the 2G has been exposed to high temperatures, but the site and modalities of the consequent nuclear processing are still debated.  No models are fully consistent with the formation of multiple populations \citep[see, e.g.][]{gratton2019}.  Collecting data and analysing the cluster-to-cluster differences in the chemical patterns helps to discriminate among models and collect information on the formation process of these fundamental galaxy constituents. \\

NGC\,6402 also hosts ``multiple populations", as shown by  \cite{johnson2019}, who determined the chemical composition of 41 NGC\,6402 giants by means of high resolution spectra obtained with the Magellan-M2FS instrument.  The \cite{johnson2019} analysis highlighted interesting peculiarities in the abundance patterns of light elements,  which prompted the present investigation. The giants in the sample were shown to belong either to the P1 (1G), 12/35 giants $\sim$34\%, or to a very mildly polluted population (P2, which here we will call 2G mild\footnote{The \cite{johnson2019} work follows the nomenclature by \cite{carretta2018}, who subdivides the stars in NGC\,2808 into five groups with increasing anomalies: P1, P2, I1, I2 and E. Roughly these groups correspond to the ChM groups B (P1), C (P2), D (I1 and I2) and E (E) by \cite{milone2015} ---see Sect.\,\ref{sec:2808comp}. Here we will use the term 1G (P1), 2G mild
(P2) and 2G extreme or E. Further nomenclature will be added in Sect.\,\ref{sec:CMDs}.
}, 14/35 giants $\sim$40\%), or to a population with `extreme' anomalies (E, 9/35 giants $\sim$26\%). The cluster is apparently lacking giants with `intermediate' (I) compositions between the 2G mild and 2G extreme or E group, and this may be due to the precise modalities of formation. The gap in chemistry is a further indication that the formation of MPs occurs discontinuously, in bursts. It may also suggest that different polluters are at play in forming the gas of the E and 2G mild populations. In fact \cite{johnson2019} propose that the E group is born in the gas processed in supermassive stars  \citep[hereinafter SMS,][]{denissenkov2014}, formed in the early phases of cluster formation by merging of massive stars, and that the P2 (2G mild) group is formed much later on, by AGB winds very diluted with re-accreting pristine gas, following the model first proposed by \cite{dercole2008};  see also \cite{calura2019} for recent simulations. \\

We refer to \cite{gratton2019} for a detailed summary of models proposed in the literature, and limit a more detailed analysis to formation of the second population in matter polluted by SMS \citep{denissenkov2014, denissenkov2015, gieles2018} and to the AGB model \citep[see, e.g.][]{dercole2008, bekki2011, dantona2016, bekki2017rprocess, ventura2018mgisotopes}. These two models have qualitatively survived a preliminary scrutiny, although with well recognized difficulty. The main problems still hampering their validation are: {\it i) for the AGB model}, the issue of the oxygen and sodium abundances in the extreme stars; {\it ii) for the SMS model}, the formation itself of such objects, and the necessity of freezing their H--burning stage to incomplete  core hydrogen burning. Consequently, observations which may help to understand whether two polluting sources are needed in the complex formation of GCs, or at least for some of them, is an issue which deserves attention and a deep analysis.

Recently attention has been also focused on the possible paths of  formation of iron--complex clusters, either along the same paths of formation  of the MPs \citep{dantona2016, lacchin2021,wirth2021}, or by considering the possible merging between clusters differing in metallicity \citep{vandenbergh1996, amaroseoane2013, bekki2016, gavagnin2016, khoperskov2018}. As NGC\,6402 will result to be a Type II cluster, we will also ask whether this feature offers clues for the formation model. 

\begin{figure*}
\centering
\includegraphics[width=9.5cm,trim={0.7cm 5cm 0.2cm 4.7cm},clip]{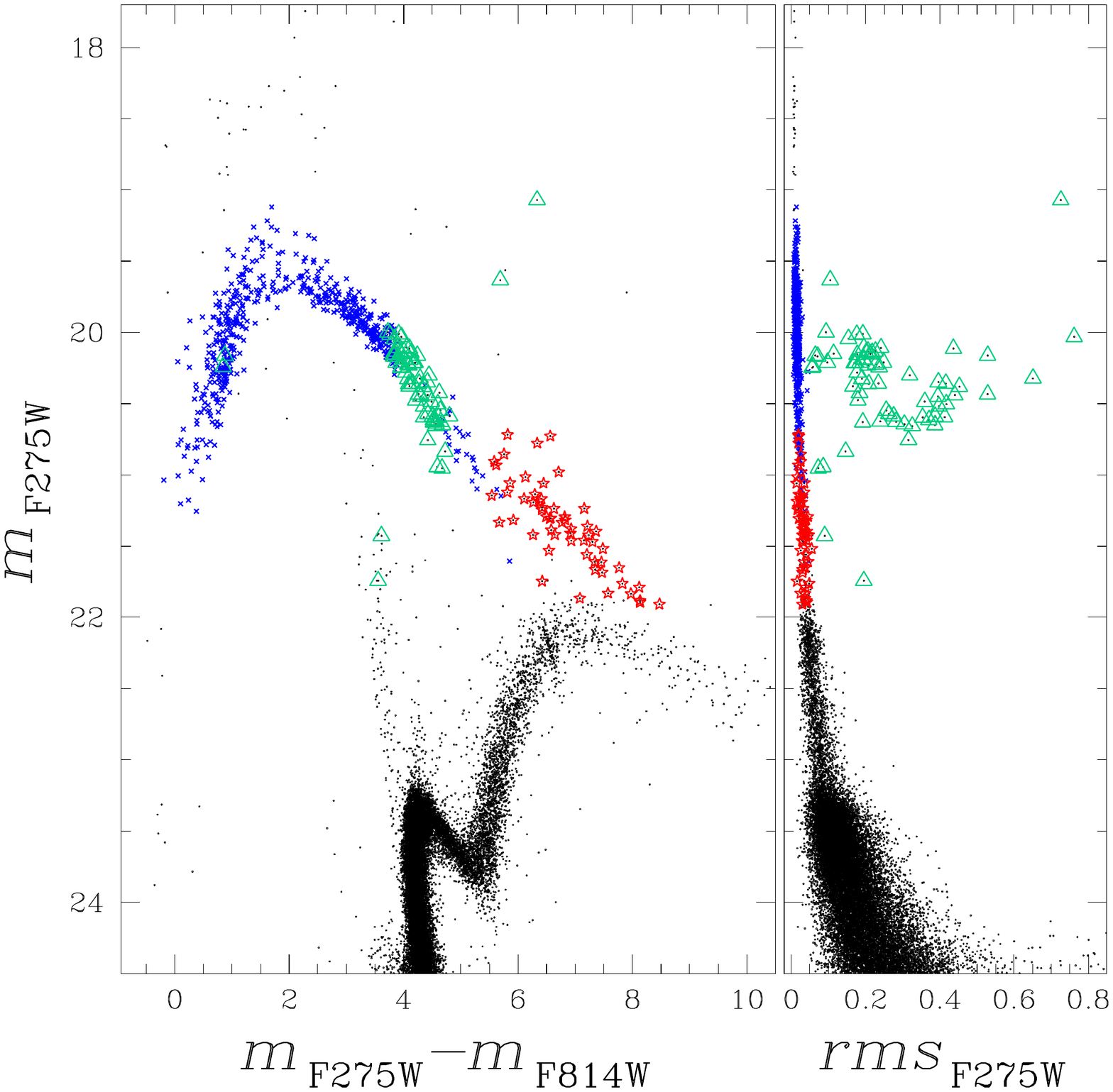}
\caption{\textit{Left Panel.} $m_{\rm F275W}$ vs.\,$m_{\rm F275W}-m_{\rm F814W}$ CMD of NGC\,6402 for stars in the central field. \textit{Right Panel.} $m_{\rm F275W}$ against the rms of the $m_{\rm F275W}$ measurements. HB non-variable stars, AGB stars, and variable stars are colored blue, red, and aqua, respectively. }
\label{fig:cmd}
\end{figure*}

\begin{figure*}
\centering
\includegraphics[width=8.5cm,trim={0.7cm 5cm 0.2cm 4.7cm},clip]{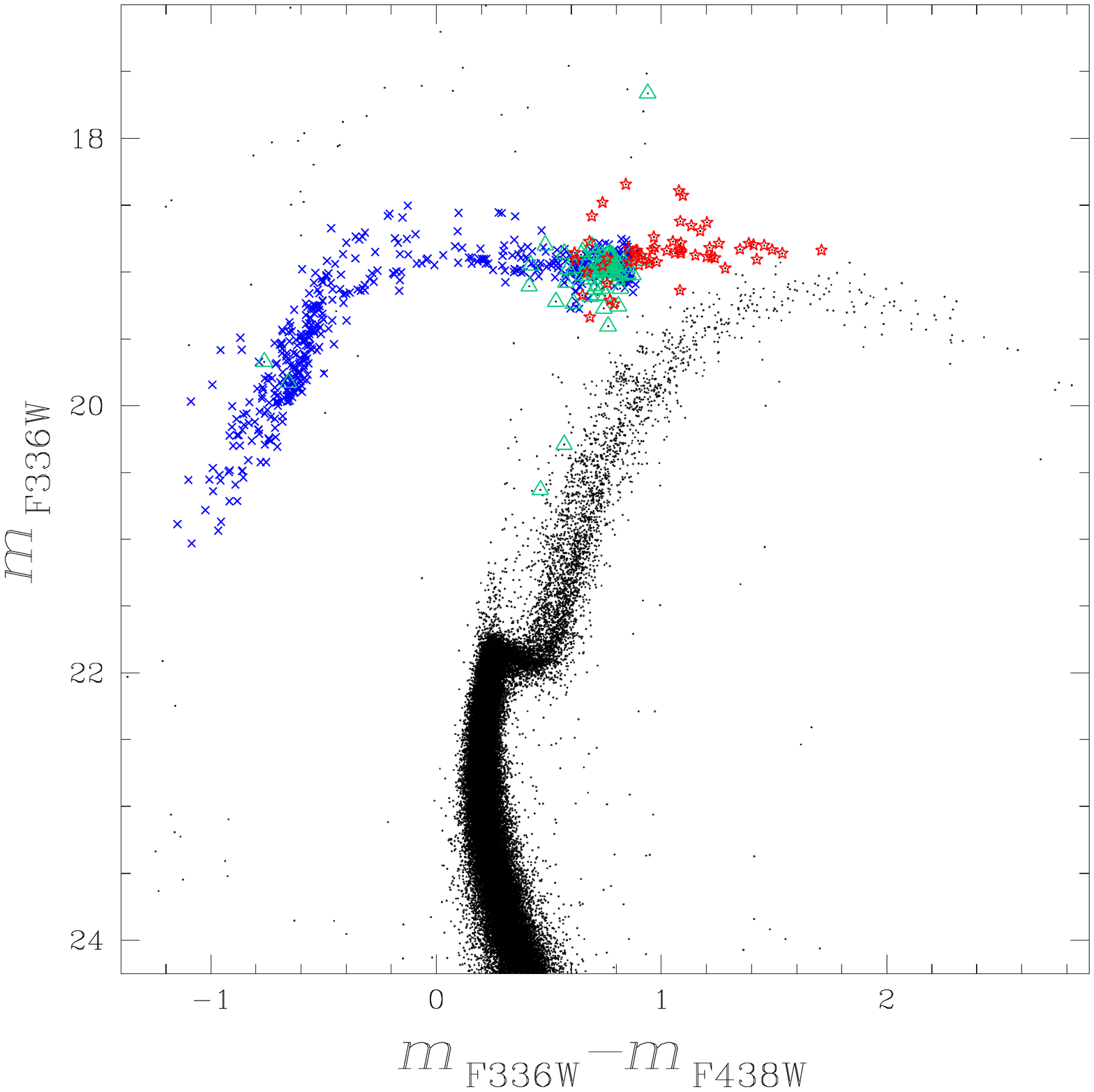}
\includegraphics[width=8.5cm,trim={0.7cm 5cm 0.2cm 4.7cm},clip]{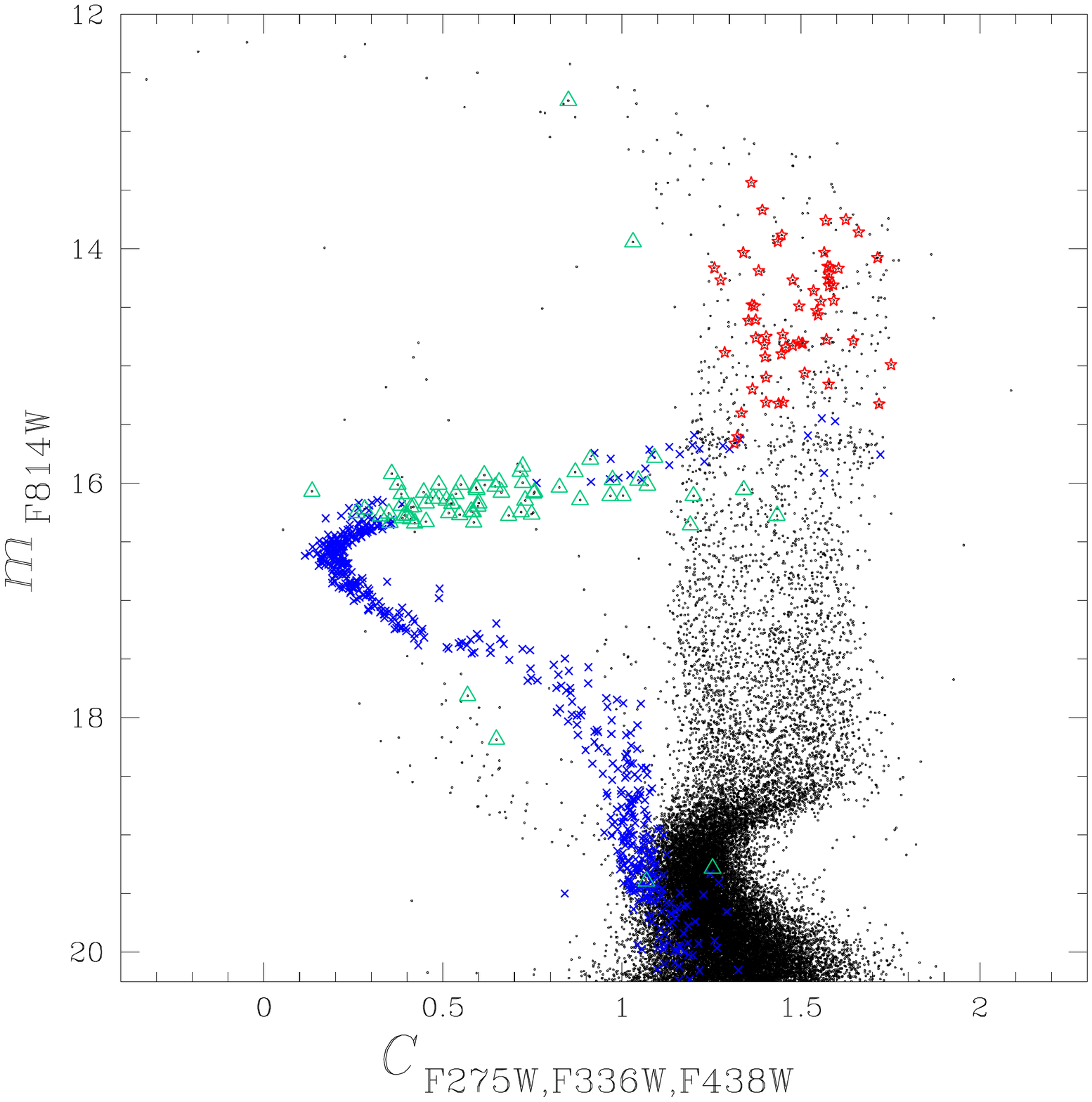}
\caption{$m_{\rm F336W}$ vs.\,$m_{\rm F336W}-m_{\rm F438W}$ CMD (left) and $m_{\rm F814W}$ vs.\,$C_{\rm F275W,F336W,F438W}$ pseudo CMD for stars in the central field. Colors are the same as in Figure\ref{fig:cmd} }
\label{fig:cmds}
\end{figure*}

The {\it HST} UV observations of the central regions of NGC\,6402 were planned to further constrain the models, by adding the photometric information useful to identify different stellar populations and combining this information with the spectroscopic analysis. In this work we present the results of the UV photometry of the core region of the cluster, its color magnitude diagrams, and the ChM, and compare them with the most similar cluster in terms of mass and metallicity, NGC\,2808. We also present and compare with NGC\,2808 the observations of a deep color magnitude diagram in the bands F475W and F814W obtained in  parallel field exposures, and give a preliminary analysis of the HB stellar distribution.\\ 

We show that NGC\,6402 hosts less extreme populations than NGC\,2808, as it could also be evinced from the lack of extreme Mg depletion in the oxygen poor giants.  The maximum helium abundance in the E population results to be Y$\sim$0.31, much smaller than the abundance required by present day massive AGB models and by the SMS models so far analysed to be consistent with strong oxygen reduction in the abundances. We briefly discuss the consequences of these results.

\section{Observations and data reduction}
To investigate multiple stellar populations we exploit {\it Hubble Space Telescope} ({\it HST}\,) images of two fields of view in the direction of NGC\,6402  
 as part of GO-16283 (PI F.\,D'Antona).
The primary field includes the cluster center and has been observed through the F275W, F336W, F438W and F814W filters of the Ultraviolet and Visual Channel of the Wide Field Camera 3 (UVIS/WFC3).
Parallel observations have been conducted with the Wide Field Channel of the Advanced Camera for Surveys (WFC/ACS) through the F475W and F814W bands. Additional information on the dataset is provided in Table \ref{tab:data}.

Photometry and astrometry was carried out with the computer program KS2 by using images corrected for the effects of poor charge transfer efficiency of the UVIS/WFC3 and WFC/ACS detectors \citep[][]{anderson2010a}. The KS2 program was written by Jay Anderson, and is the evolution of $kitchen\_sink$, originally developed to reduce two-filter ACS/WFC images \citep{anderson2008}.
KS2 follows different recipes to derive stellar fluxes and positions. The first method measures the stars in each exposure independently, by fitting the appropriate effective point-spread function (PSF) models. The various measurements are then averaged together to get the best estimates of magnitudes and positions.
 This method provides the best measurements of bright stars. 
 To measure faint stars that have not enough flux to be properly constrained by the PSF fit, KS2 combines the information from all exposures by fixing the average stellar positions from all exposures. 
 After subtracting neighbor stars, it measures the faint stars by means of aperture photometry. We refer to papers by \citet{sabbi2016a} and \citet{nardiello2018a} for details on KS2. \\
 
 Stellar positions have been corrected for geometric distortion by using the solutions presented in \citet{bellini2009a} and \citet{bellini2011a}. Photometry has been calibrated to the Vega magnitude system by using the zero points of the WFC/ACS and UVIS/WFC3 filters available at the Space Telescope Science Institute web pages.
 Since we are interested in investigating multiple populations, we selected the sample of well measured stars, which are relatively isolated and well fitted by the PSF \citep[see][for details]{milone2009a}. To do this, we exploited the diagnostics provided by the KS2 program, including magnitude rms, the fraction of flux in the aperture due to neighbours and the quality of the PSF fit.  Finally, the photometry was corrected for differential reddening by using the method and computer program described in \citet{milone2012}.

\begin{table}
  \caption{Description of the {\it HST} images used in the paper. All images are collected on February 6$^{\rm th}$-13$^{\rm rd}$ 2021 as part of GO-16283 (PI F.\,D'Antona).}

\begin{tabular}{c c c }
\hline \hline
 CAMERA & FILTER  & N$\times$EXPTIME \\
\hline
 & Primary Field & \\
\hline
 UVIS/WFC3 & F275W & 987s$+$988s$+$10$\times$1050s$+$     \\
 &  & 3$\times$1172s+1181s     \\
 UVIS/WFC3                  & F336W & 11$\times$761s \\
 UVIS/WFC3                  & F438W & 2$\times$100s+10$\times$348s \\
 UVIS/WFC3                  & F814W & 10$\times$110s \\
\hline
 & Parallel Field & \\
\hline 
  WFC/ACS & F475W &  350s$+$822s$+$2$\times$890s$+$ \\
  & & 10$\times$990s$+$1046s$+$1091s    \\
  WFC/ACS & F814W & 35s$+$350s$+$10$\times$619s$+$719s\\
     \hline\hline
\end{tabular}
  \label{tab:data}
 \end{table}

\section{Photometric diagrams of NGC 6402}
\label{sec:CMDs}
The differential-reddening corrected photometry has been used to build photometric diagrams that are sensitive to multiple stellar populations. 

\subsection{The Color Magnitude diagrams}
\label{sec:cmd}
The left panel of Figure \ref{fig:cmd} shows $m_{\rm F275W}$ vs.\,$m_{\rm F275W}-m_{\rm F814W}$ for stars in the central field. Thanks to the wide color baseline, this CMD maximizes the effective-temperature differences among MS and RGB stars with similar luminosities and provides an exquisite tool to identify stellar populations with different helium abundances. 
The right panel of Figure \ref{fig:cmd} shows the rms of the F275W magnitude determinations against $m_{\rm F275W}$. The fact that some stars exhibit large rms values, when compared with the bulk of stars of similar magnitude, is a signature of stellar variability.
The sample of candidate variables comprises 71 stars that are represented with aqua triangles in both panels of Figure \ref{fig:cmd}, including 65 candidate RR Lyrae, two blue stragglers, two hot-HB stars, and two post-HB stars. \\


Two additional CMDs are provided in Figure \ref{fig:cmds} for stars in the central field. A split RGB is clearly visible in the $m_{\rm F336W}$ vs.\,$m_{\rm F336W}-m_{\rm F438W}$ CMD, which is the {\it HST} analogous of $U$ vs.\,$U-B$ and is mostly sensitive to nitrogen variations \citep[][]{marino2008a}.  The $m_{\rm F814W}$ vs.\,$C_{\rm F275W,F336W,F438W}$ also reveals a broad RGB, thus confirming that NGC\,6402 hosts stellar populations with different light-element abundances. Intriguingly, the AGB of NGC\,6402 exhibits a wide $C_{\rm F275W,F336W,F438W}$ intrinsic spread, which is comparable with that of RGB stars with similar luminosity. This fact demonstrates that, similarly to the RGB, the AGB of NGC\,6402 hosts multiple populations.  

\begin{figure}
\centering
\includegraphics[width=9.0cm,trim={0.7cm 5.5cm 0.2cm 14.cm},clip]{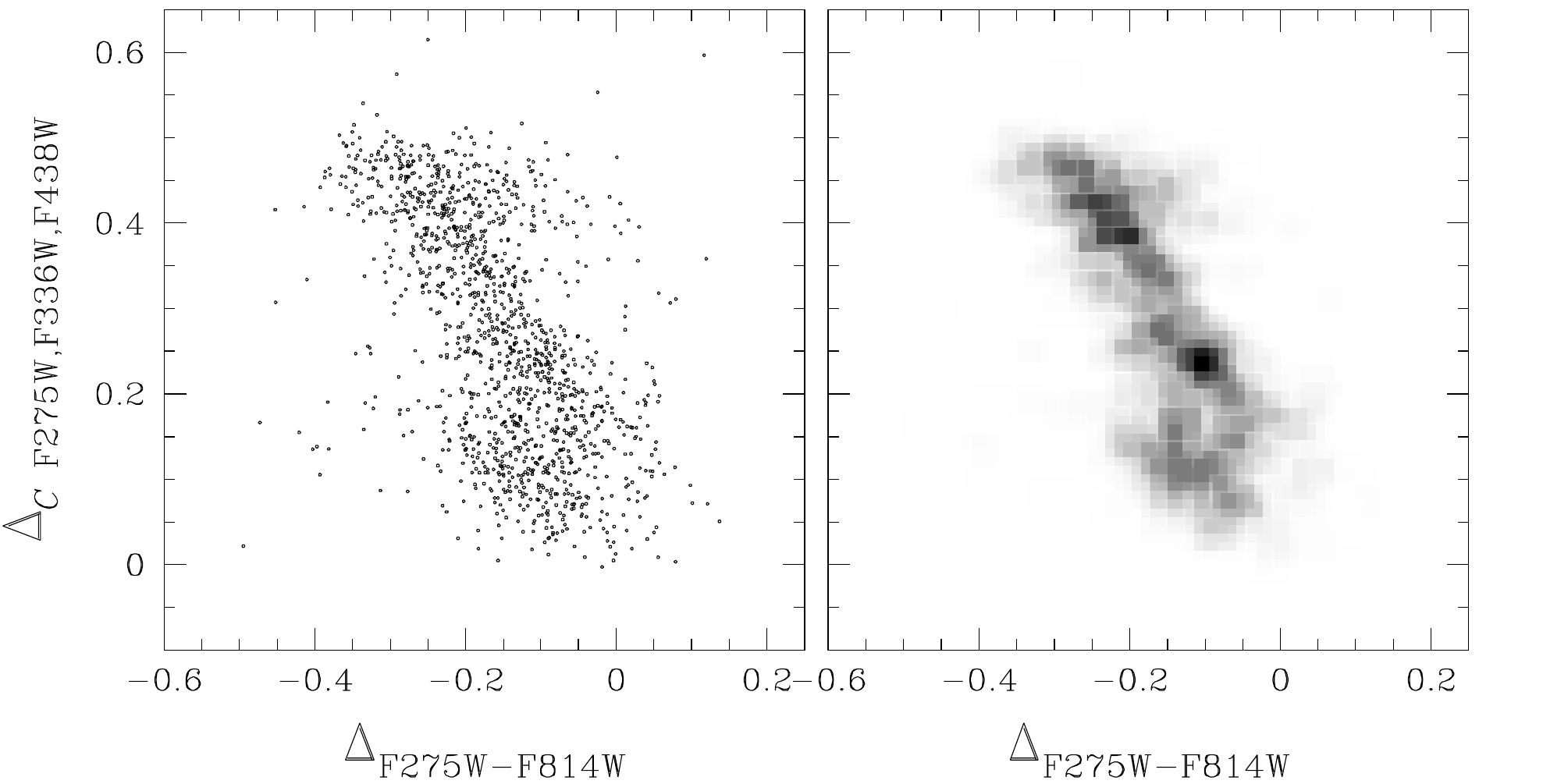}
\includegraphics[width=9.0cm,trim={0.7cm 5.5cm 0.2cm 14.cm},clip]{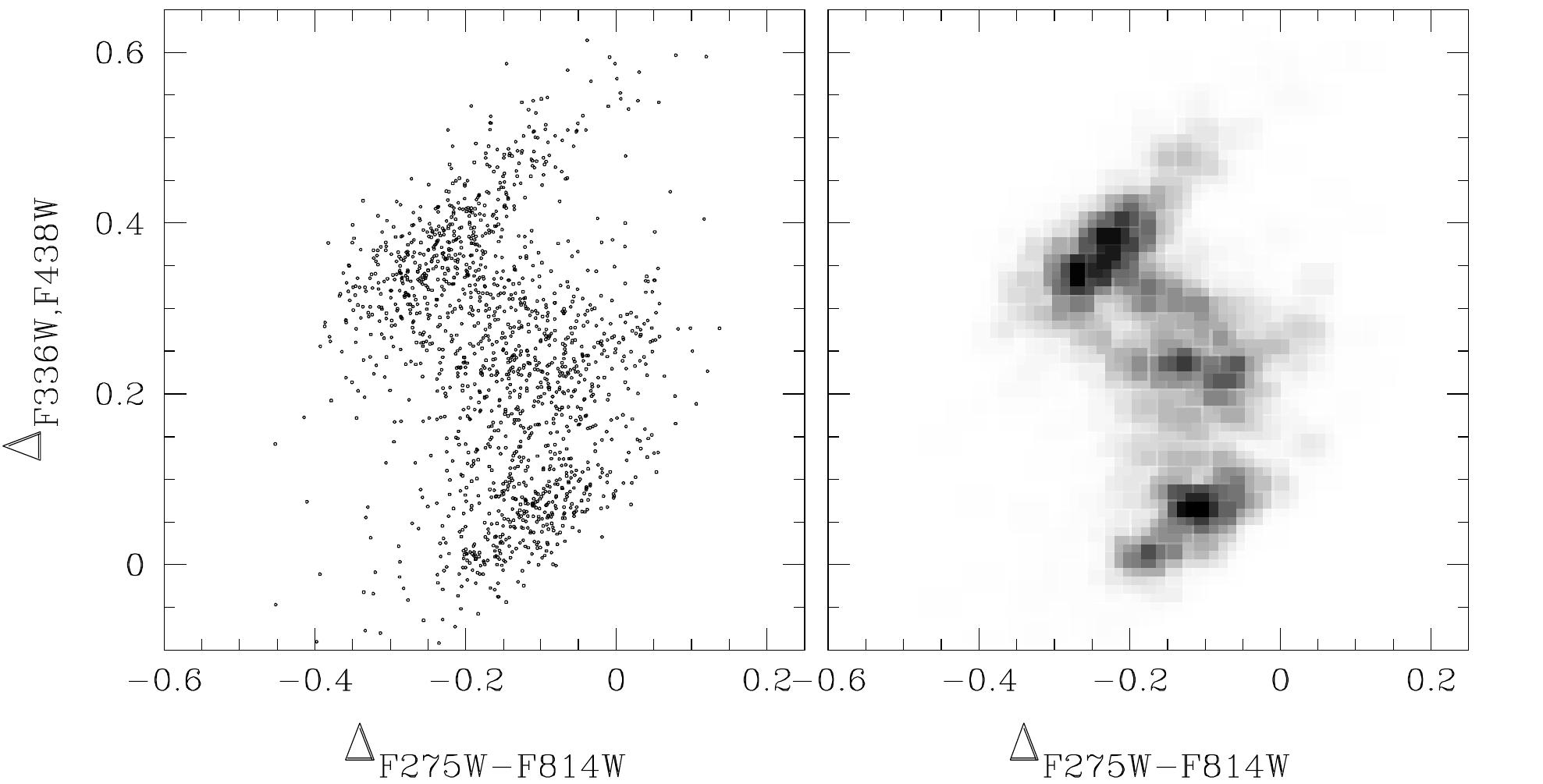}
\caption{Left panels show $\Delta_{C, {\rm F275W,F336W,F438W}}$ vs.\,$\Delta_{\rm F275W-F814W}$ (top) and  $\Delta_{\rm F336W-F438W}$ vs.\,$\Delta_{\rm F275W-F814W}$ ChMs (bottom) of RGB stars in NGC\,6402. The corresponding Hess diagrams are plotted in the right panels.}
\label{fig:ChMs}
\end{figure}

\subsection{The Chromosome map}
\label{sec:chm}
The Chromosome map (ChM) is a pseudo two-color diagram of MS, RGB, or AGB stars obtained from appropriate filter combinations that maximize the separation among stellar populations in GCs \citep[][]{milone2015, marino2017a}.  
However, the ChM differs from a simple two-color diagram because the sequence of stars is verticalized on both colors.  \\

\begin{figure*}
\centering
\includegraphics[height=5.5cm,trim={0.7cm 5.5cm 9.5cm 13.75cm},clip]{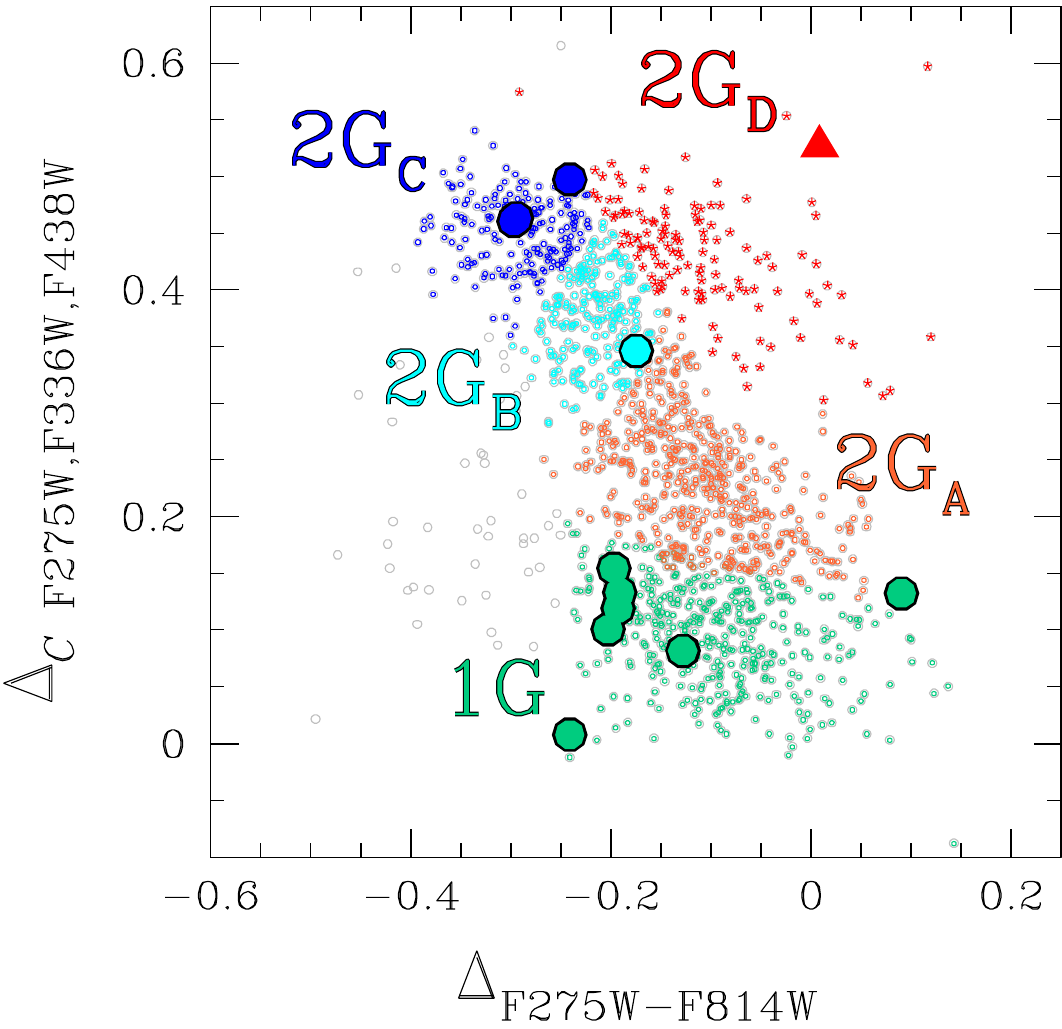}
\includegraphics[height=5.84cm,trim={0.7cm 5.5cm 0.2cm 13.75cm},clip]{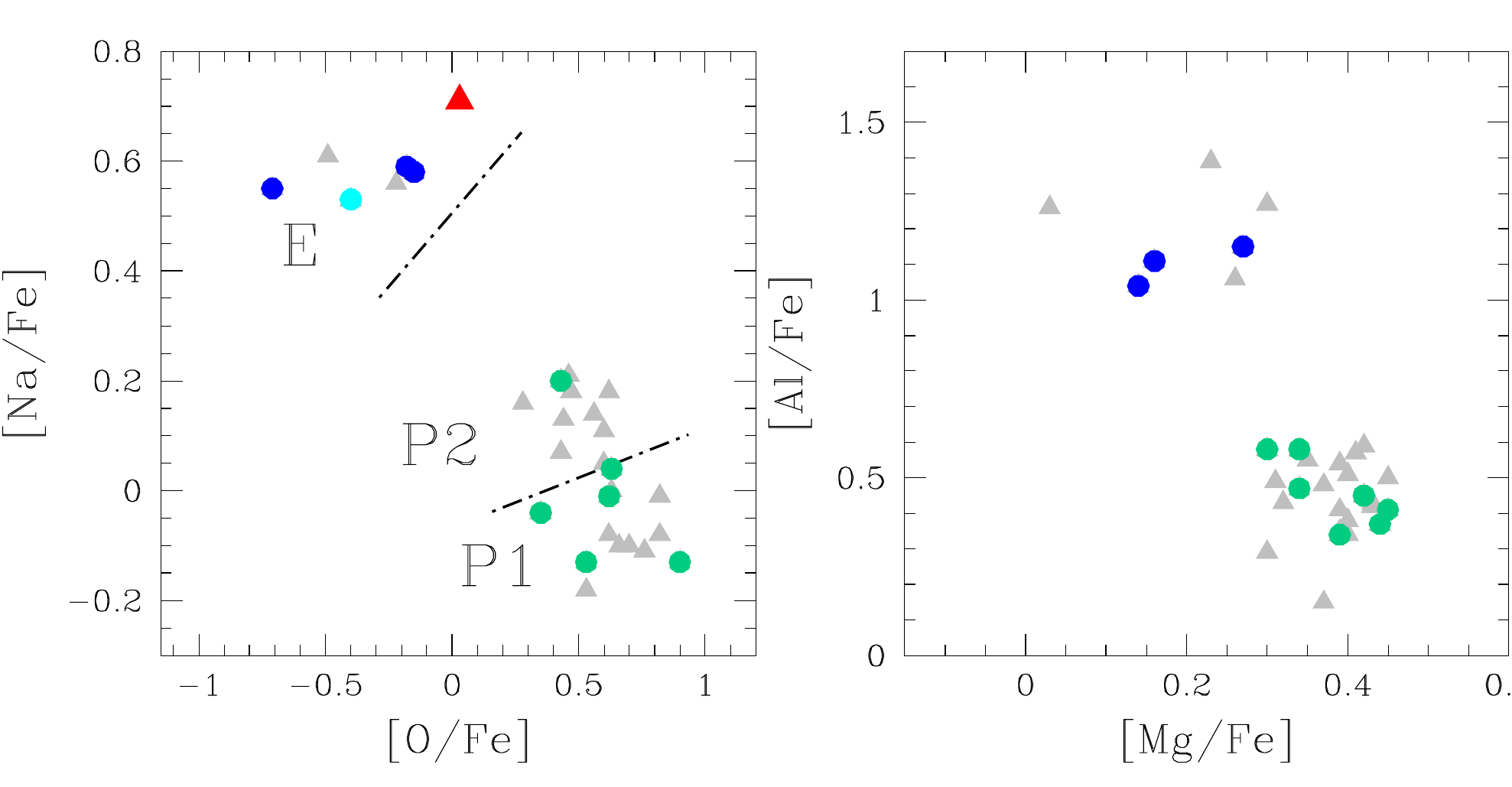}
\caption{Reproduction of the $\Delta_{C, {\rm F275W,F336W,F438W}}$ vs.\,$\Delta_{\rm F275W-F814W}$ ChM of Figure \ref{fig:ChMs} (Left panel). Middle and left panel show the sodium-oxygen  and the magnesium-aluminum anticorrelations, respectively, from \citet{johnson2019}. 
The dashed-dotted lines in the middle panel separate the populations of P1, P2 and E stars defined by Johnson and collaborators.
1G, 2G$_{\rm A}$, 2G$_{\rm B}$, 2G$_{\rm C}$ and 2G$_{\rm D}$ stars, selected from the ChM, are colored aqua, orange, cyan, blue, and red, respectively. Stars for which both photometry and spectroscopy are available are represented with large colored  symbols.}
\label{fig:chimica}
\end{figure*}

Here, we derived two distinct ChMs of RGB stars based on the $m_{\rm F814W}$ vs.\,$m_{\rm F275W}-m_{\rm F814W}$, the $m_{\rm F814W}$ vs.\,$m_{\rm F336W}-m_{\rm F438W}$ CMD, and the $m_{\rm F814W}$ vs.\,$C_{\rm F275W,F336W,F438W}$ pseudo CMD by following the recipe by 
 \citet[][see their Section 3.1]{milone2015, milone2017chromo}\footnote{In a nutshell, to derive the $\Delta_{\rm F275W,F814W}$ pseudo color of RGB stars, we divided the $m_{\rm F814W}$ vs.\,$m_{\rm F275W}-m_{\rm F814W}$ CMD into 0.1-mag wide magnitude bins and calculated the 4$^{\rm th}$ and the 96$^{\rm th}$ percentiles of the $m_{\rm F275W}-m_{\rm F814W}$ color distributions of RGB stars in each bin. These quantities have been associated with the median values of the F814W magnitude distribution of the stars in each bin and have been linearly interpolated to derive the blue and red boundaries of the RGB. The $\Delta_{\rm F275W,F814W}$ pseudo-color has been derived by means of Equation 1 from \citet{milone2017chromo}, which tranforms the CMD into a 'verticalized' diagram, where the blue and red RGB boundaries translate into vertical lines with $\Delta_{\rm F275W,F814W}$=0 and $\Delta_{\rm F275W,F814W}$=$-W_{\rm F275W,F814W}$, where $W_{\rm F275W,F814W}=0.34$ mag is the RGB width calculated 2.0 F814W magnitudes above the MS turn off.  A similar approach has been adopted to derive the $\Delta_{C \rm F275W,F336W,F438W}$ pseudo color but by using $m_{\rm F814W}$ vs.\,$C_{\rm F275W,F336W,F438W}$ pseudo CMD. In this case, the diagram has been verticalized by assuming Equation 2 from \citet{milone2017chromo} and the $C_{\rm F275W,F336W,F438W}$ RGB width of $W_{\rm F275W,F336W,F438W}=0.48$ mag. See \citet[][]{milone2017chromo} for details.
}.
The $\Delta_{\rm {\it C} F275W,F336W,F438W}$ vs.\,$\Delta_{\rm F275W-F814W}$ ChM of RGB stars and the corresponding Hess diagram are shown in the upper panels of Figure \ref{fig:ChMs}, whereas in the bottom panels we plotted $\Delta_{\rm F336W-F438W}$  against $\Delta_{\rm F275W-F814W}$. 
The ChM is shown again in the left panel of Fig.\ref{fig:chimica}. It highlights the group of 1G stars, located around the origin of the map and colored in green, 
and an extended 2G, where we identified four sub-populations of 2G$_{\rm A}$, 2G$_{\rm B}$, 2G$_{\rm C}$ and 2G$_{\rm D}$ stars. Stars that are likely members of these groups are colored orange, cyan, blue, and red, respectively. NGC\,6402 is affected by a large average reddening \citep[E(B-V)=0.6,][]{harris1996} and by significant reddening variations across the field of view.
To verify that the multiple populations identified on the ChM and the CMD are not artefacts due to differential reddening,  we checked that each population is distributed in the entire field of view. Moreover, we divided the internal field into 39 quadrants and  analysed the ChM of stars in each of them. All the five stellar populations identified in Fig.\,\ref{fig:chimica} are present in the ChM of stars in each quadrant, proving that they are real features of NGC\,6402.\\

2G$_{\rm D}$ stars define a tail of stars in the ChM with redder $\Delta_{\rm F275W-F814W}$ values than the bulk of stars with similar $\Delta_{C, {\rm F275W,F336W,F438W}}$ pseudo-colors.
This region of the ChM is typically populated by stars with enhanced iron and/or C$+$N$+$O abundance, and is a distinctive feature of Type\,II GCs \citep{milone2017chromo}. \\

By extending  the procedure by \citet[][see their Section 3.3]{zennaro2019a} to the ChM plotted in Figure\,\ref{fig:chimica}, we find that the 1G hosts 27.9$\pm$2.3\% of the studied RGB stars. The selected groups of 2G stars include 34.6$\pm$2.4\% (2G$_{\rm A}$), 16.1$\pm$2.0\% (2G$_{\rm B}$), 12.9$\pm$1.4\% (2G$_{\rm C}$) and 8.4$\pm$1.3\% (2G$_{\rm D}$).  \\

To infer the helium content of stellar populations, we adopted for NGC\,6402 the method widely used by our group and used to constrain helium variations in about 60 GCs \citep[e.g.][]{milone2018he, lagioia2019a}. In a nutshell, we derived the RGB fiducial lines of each population in the $m_{\rm F814W}$ vs.\,$m_{\rm X}-m_{\rm F814W}$, where X=F275W, F336W, and F438W. We defined five equally-spaced F814W magnitude values in the interval between $m_{\rm F814W}=14.8$ and 18.0. For each value we calculated the $m_{\rm X}-m_{\rm F814W}$ color difference between each fiducial and the 1G ones. The comparison between the observed colors and appropriate grid of synthetic spectra with different abundances of He, C, N and O provides an estimate of the relative abundances of these elements. \\

Results are listed in Table \ref{tab:He} and show that group 2G$_{\rm C}$ is highly helium enhanced, by $\Delta$Y$\sim 0.05$\ with respect to the 1G and the 2G$_{\rm A}$, which are both assumed to have pristine helium abundance (Y$\sim0.25$). 

Both 2G$_{\rm B}$ and 2G$_{\rm C}$ are enriched in nitrogen with respect to the 1G by more than 1 dex and depleted in carbon and oxygen by $\sim$0.4 and 0.8 dex, respectively. 
On the contrary,  population 2G$_{\rm A}$ shares similar C and O abundances as the 1G but is enriched in [N/Fe] by $\sim$0.6 dex. 
We did not infer the chemical composition of 2G$_{\rm D}$\  stars, due to the enhanced metallicity of these stars. Since variations in iron significantly affect the stellar colors (and it would be challenging to disentangle the effect of helium and iron), we are not able to provide robust determinations of both helium and metallicity for these stars. Nevertheless, given the extreme  $\Delta_{C, {\rm F275W,F336W,F438W}}$\ pseudo-color, it would be reasonable to suggest that these 2G$_{\rm D}$\ stars have extreme chemical composition, hence high helium and nitrogen content, and are depleted in oxygen and carbon. If this hypothesis is correct,  2G$_{\rm D}$ stars are the RGB counterparts of a fraction of blue-MS stars.
\\
\begin{table}
\centering
  \caption{Chemical composition of stellar populations relative to 1G stars. We assumed all population  [Fe/H]=$-1.10$ and adopted for the 1G Y=0.25, solar C and N content and [O/Fe]=0.3. }

\begin{tabular}{c c c c c}
\hline \hline
  Pop.  & $\Delta$Y  & $\Delta$[C/Fe]  &  $\Delta$[N/Fe]  & $\Delta$[O/Fe]  \\  
  \hline
  2G$_{\rm A}$ & 0.000$\pm$0.010 &    0.00$\pm$0.20 & 0.60$\pm$0.15 & $-$0.10$\pm$0.10 \\
  2G$_{\rm B}$ & 0.037$\pm$0.007 & $-$0.40$\pm$0.20 & 1.10$\pm$0.10 & $-$0.80$\pm$0.15 \\
  2G$_{\rm C}$ & 0.051$\pm$0.009 & $-$0.40$\pm$0.20 & 1.20$\pm$0.15 & $-$0.80$\pm$0.15 \\
   \hline\hline
\end{tabular}
  \label{tab:He}
 \end{table}

\subsubsection{Comparison with Johnson et al.\,(2019).}
As discussed in the introduction, the first evidence of multiple populations in NGC\,6402 was provided by \citet[][]{johnson2019}, who analyzed 41 giant stars by using  high resolution spectra collected with Magellan-M2FS. 
They inferred chemical abundances of eleven elements and identified three main stellar populations based on their position in the [Na/Fe] vs.\,[O/Fe] plane. The middle panel of Figure \ref{fig:chimica} reproduces the sodium-oxygen anti-correlation by Johnson and collaborators and highlights the three populations of P1, P2, and E stars identified by these authors. \\

To further constrain the chemical composition of stellar populations identified along the ChM, we exploit eleven stars for which both photometry and spectroscopy from \citet[][]{johnson2019} are available (large colored symbols in Figure \ref{fig:chimica}). \\

Results are listed in Table \ref{tab:abbJ19} where we show for each population with available spectroscopic targets, the average elemental abundance, the random mean scatter and the number of spectroscopic targets. 
We find that 1G stars have nearly solar sodium content and are enhanced in both oxygen and magnesium. On the contrary, 2G$_{\rm C}$ are depleted in both [O/Fe] and [Mg/Fe] and enhanced in [Na/Fe], with respect to the 1G. \\

The comparison between spectroscopic and photometric results provides the opportunity to associate the populations identified in this paper and by \citet[][]{johnson2019}. Clearly the 1G corresponds to the population P1 by Johnson and collaborators, while their population E is composed of stars in the groups 2G$_{\rm B}$, 2G$_{\rm C}$ and 2G$_{\rm D}$.  \\

It would be tempting to associate the 2G$_{\rm A}$\ stars identified on the ChM with the P2 by \citet[][]{johnson2019}, as they exhibit moderate enhancement in N and Na, respectively, with respect to the 1G.  Unfortunately there are no elemental abundances for 2G$_{\rm A}$ stars.
 Moreover, the only P2 star with available photometry seems located on the 1G of the ChM despite it is sodium enhanced by more than 0.2 dex with respect to the bulk of 1G stars. 
 More data are needed to establish whether the  2G$_{\rm A}$\  is composed of 1G stars or if the analysed P2 stars has large photometric and/or spectroscopic uncertainties.

\begin{table}
\centering
  \caption{Chemical composition of 1G, 2G$_{\rm B}$, 2G$_{\rm C}$ and 2G$_{\rm D}$ stars of NGC\,6402 inferred from the chemical abundances derived by \citet[][]{johnson2019}.}

\begin{tabular}{c c c c c}
\hline \hline
  & Population & mean  & r.m.s  & N \\  
  \hline
$\rm [O/Fe]$  &   1G              &   0.58  & 0.19  & 6 \\
              &   2G$_{\rm B}$    &$-$0.40  & ---   & 1 \\
              &   2G$_{\rm C}$    &$-$0.35  & 0.32  & 3 \\
              &   2G$_{\rm D}$    &   0.03  & ---   & 1 \\
\hline              
$\rm [Mg/Fe]$ &   1G              &   0.38  & 0.06  & 7 \\
              &   2G$_{\rm C}$    &   0.19  & 0.07  & 3 \\
\hline
$\rm [Al/Fe]$ &   1G              &   0.45  & 0.06  & 7 \\
              &   2G$_{\rm B}$    &   1.27  & ---   & 1 \\
              &   2G$_{\rm C}$    &   1.10  & 0.06  & 3 \\
              &   2G$_{\rm D}$    &   0.90  & ---   & 1 \\
\hline
$\rm [Si/Fe]$ &   1G              &   0.30  & 0.08  & 7 \\
              &   2G$_{\rm B}$    &   0.35  & ---   & 1 \\
              &   2G$_{\rm C}$    &   0.38  & 0.02  & 3 \\
              &   2G$_{\rm D}$    &   0.66  & ---   & 1 \\
\hline
$\rm [Na/Fe]$ &   1G              &$-$0.01  & 0.11  & 7 \\
              &   2G$_{\rm B}$    &   0.53  & ---   & 1 \\
              &   2G$_{\rm C}$    &   0.57  & 0.02  & 3 \\
              &   2G$_{\rm D}$    &   0.71  & ---   & 1 \\
\hline
$\rm [Ca/Fe]$ &   1G              &   0.29  & 0.05  & 7 \\
              &   2G$_{\rm B}$    &   0.42  & ---   & 1 \\
              &   2G$_{\rm C}$    &   0.38  & 0.01  & 3 \\
              &   2G$_{\rm D}$    &   0.33  & ---   & 1 \\
\hline
$\rm [Fe/H]_{I}$  &   1G              &$-$1.14  & 0.04  & 7 \\
              &   2G$_{\rm B}$    &$-$1.10  & ---   & 1 \\
              &   2G$_{\rm C}$    &$-$1.14  & 0.04  & 3 \\
              &   2G$_{\rm D}$    &$-$1.05  & ---   & 1 \\
\hline
$\rm [Fe/H]_{II}$  &   1G              &$-$1.15  & 0.04  & 7 \\
              &   2G$_{\rm B}$    &$-$1.11  & ---   & 1 \\
              &   2G$_{\rm C}$    &$-$1.14  & 0.04  & 3 \\
              &   2G$_{\rm D}$    &$-$1.06  & ---   & 1 \\
\hline
$\rm [Cr/Fe]$ &   1G              &   0.04  & 0.07  & 7 \\
              &   2G$_{\rm B}$    &   0.13  & ---   & 1 \\
              &   2G$_{\rm C}$    &   0.13  & 0.07  & 3 \\
              &   2G$_{\rm D}$    &   0.06  & ---   & 1 \\
\hline
$\rm [La/Fe]$ &   1G              &   0.28  & 0.02  & 7 \\
              &   2G$_{\rm B}$    &   0.33  & ---   & 1 \\
              &   2G$_{\rm C}$    &   0.32  & 0.14  & 3 \\
              &   2G$_{\rm D}$    &   0.18  & ---   & 1 \\
\hline
$\rm [Eu/Fe]$ &   1G              &   0.32  & 0.06  & 6 \\
              &   2G$_{\rm B}$    &   0.36  & ---   & 1 \\
              &   2G$_{\rm C}$    &   0.41  & 0.02  & 2 \\
              &   2G$_{\rm D}$    &   0.31  & ---   & 1 \\

   \hline\hline
\end{tabular}
  \label{tab:abbJ19}
 \end{table}

\subsection{Parallel-field photometry}

\begin{figure*}
\centering
\includegraphics[width=12.5cm,trim={0.7cm 5cm 0.2cm 7.7cm},clip]{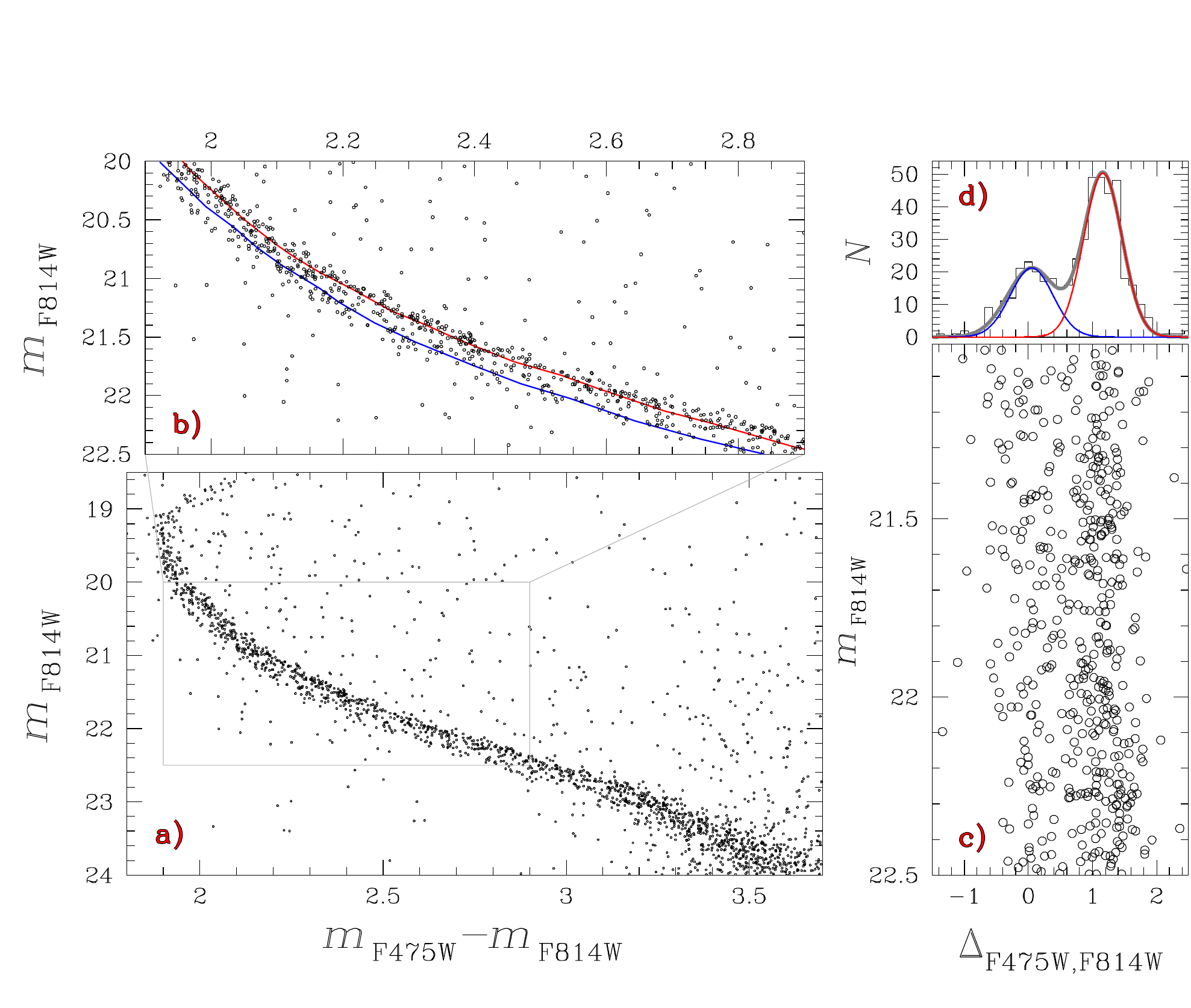}
\caption{This figure illustrates the procedure to estimate the fraction of blue- and red-MS stars. Panel a  
 shows the $m_{\rm F814W}$ vs.\,$m_{\rm F475W}-m_{\rm F814W}$ CMD of stars in the parallel field 
  while  Panel b is a zoom in of stars in the F814W interval where the split MS is more clearly visible. The red and blue lines superimposed on the CMD are the  fiducials of the corresponding MSs. The verticalized $m_{\rm F814W}$ vs.\,$\Delta (m_{\rm F475W}-m_{\rm F814W})$ diagram of panel-b stars is shown in the panel c) and the corresponding $\Delta (m_{\rm F475W}-m_{\rm F814W})$ histogram distribution is illustrated in panel d. The best fit bi-Gaussian function is superimposed on the histogram (gray thick line) and the two components are colored blue and red.}
\label{fig:pratio}
\end{figure*}

The $m_{\rm F814W}$ vs.\,$m_{\rm F475W}-m_{\rm F814W}$ CMD corrected for differential-reddening of stars in the parallel field is plotted 
 in Figure \ref{fig:pratio}a. Clearly, NGC\,6402 exhibits a split MS in the magnitude interval $20.0<m_{\rm F814W}<22.5$ and the two sequences seem to merge together along the MS turnoff and the SGB. To further demonstrate that the double MS is not due to reddening variations, we verified that red- and blue-MS stars are distributed along the entire field of view, and that the double MS is evident in the CMDs of stars of each of the nine quadrants that compose the field of view. \\

To estimate the fraction of stars in the blue and red MS we adopted the procedure illustrated in Figure \ref{fig:pratio}. In a nutshell, we selected the region of the CMD where the MS split is evident (panel a) and derived the fiducial lines of the red and blue MS (red and blue line, respectively, in panel b). These two lines are used to verticalize the MS, in such a way that the fiducial lines of the blue and the red MS translate into two vertical lines with color residuals $\Delta_{\rm F475W,F814W}$=0 and  $\Delta_{\rm F475W,F814W}$=1, respectively \citep[panel c, see][for details]{milone2015}. Finally, the $\Delta_{\rm F475W,F814W}$ histogram distribution plotted in panel d is fitted with a function provided by the sum of two Gaussian curves, by means of least squares. From the area of the two Gaussian components (red and blue curves in panel d of Figure \ref{fig:pratio}) we infer that 65.2$\pm$2.3 \% of stars belong to the red MS, while the blue MS is composed of the remaining 34.8$\pm$2.3 \% of MS stars.
\begin{figure}
    \centering
 \includegraphics[width=\columnwidth,trim={0cm 0cm 0cm 0cm}]{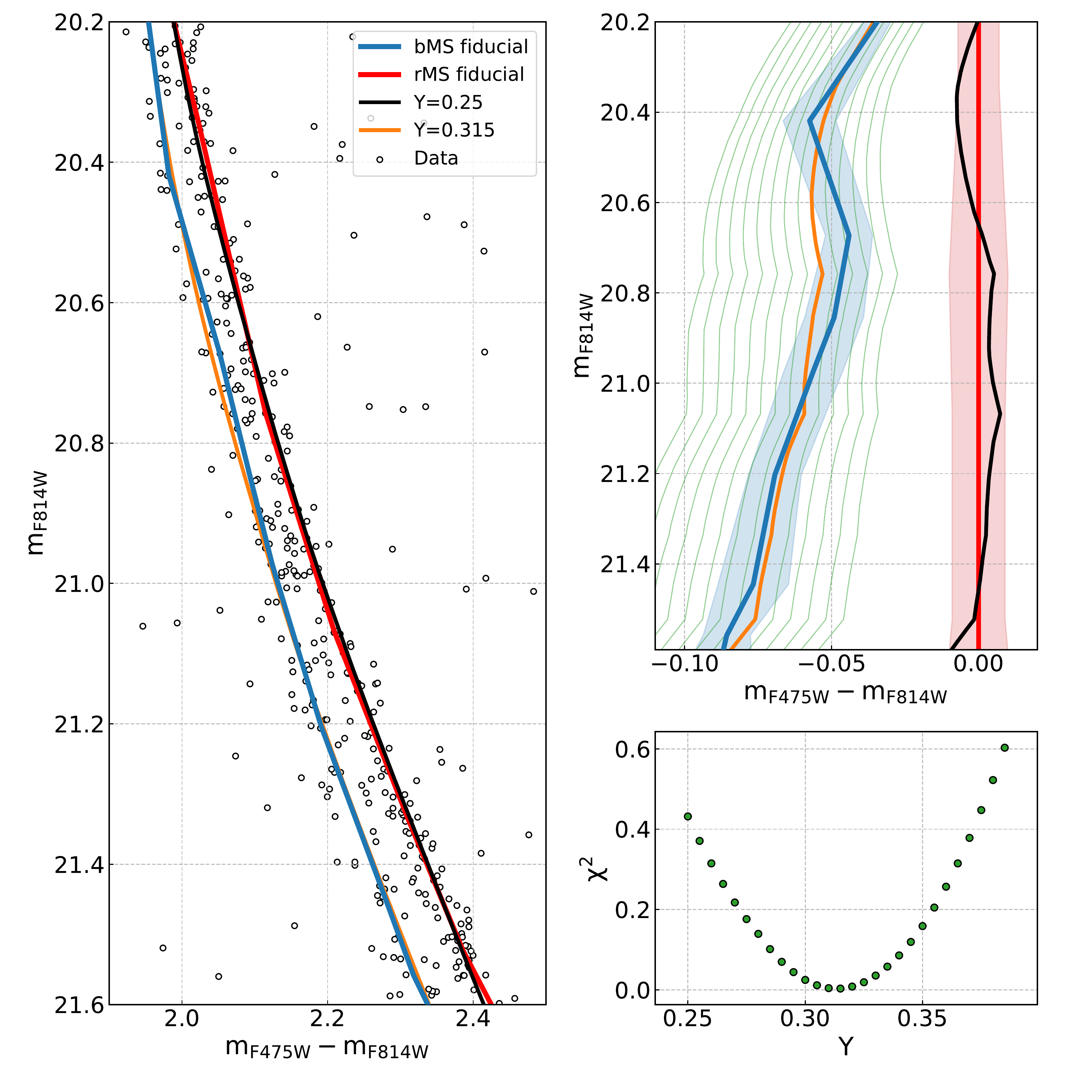}
    \caption{
    Procedure to estimate the helium abundance of the blue MS in NGC\,6402.
    Left panel shows the $m_{\rm F814W}$ vs.\,$(m_{\rm F475W}-m_{\rm F814W})$ diagram for MS stars in the parallel field. Red and blue lines represent the fiducials of the red and blue MS, respectively, while the best-fit isochrones with Y=0.25 and Y=0.315 are colored black and orange, respectively. Upper-right panel reproduces the verticalized $m_{\rm F814W}$ vs.\,$\Delta (m_{\rm F475W}-m_{\rm F814W})$ diagram. Here we added isochrones with Y ranging from Y=0.29 to Y=0.36 in steps of 0.005 (green lines). The shaded blue and red areas enclose the regions within $\pm$1-$\sigma$ color from the corresponding fiducial. Bottom-right panel shows the $\chi^{2}$ value against the helium abundance of the isochrone used to fit the blue MS. See text for details. 
    %
    }
    \label{pic:helium_2ge}
\end{figure}

In monometallic GCs the $m_{\rm F475W}-m_{\rm F814W}$ color split of MS stars is due to helium variations. To infer the average helium difference between blue-MS and red-MS stars we compared the observed MSs with appropriate isochrones from the ATON database \citep[][]{tailo2016, tailo2020} with different helium abundances by using the procedure illustrated in Figure \ref{pic:helium_2ge}. \\

We adopted for all isochrones the same iron abundance, [Fe/H]=$-1.1$, and [$\alpha$/Fe]=0.3 as in \citet[][]{johnson2019}. The isochrone with Y=0.25 that provides the best fit with the observed MSTO, SGB, and red MS is derived as in \citet[][]{tailo2020} and corresponds to age of 12.50$\pm$0.75 Gyr, distance modulus of (m$-$M)$_0$=14.76 mag,  and reddening of E(B$-$V)=0.62 mag. \\


%
%
To derive the helium abundance of the blue MS, we defined a grid of six reference magnitude values in the interval where the split MS is clearly visible. Specifically, reference magnitudes range from $m_{\rm F814W}=$20.4 to $m_{\rm F814W}=$21.4 in steps of 0.20 mag.
To infer the relative helium content of the two MSs, we first calculated the $m_{\rm F475W}-m_{\rm F814W}$  color difference between the fiducial lines of the red and blue MS corresponding to the six magnitude values. Then, we considered a grid of helium-enhanced isochrones with Y ranging from 0.250 to 0.385 in steps of 0.005.
We derived the color differences between best-fit isochrone with Y=0.25 and each helium enhanced isochrone and compared these color differences with the observed ones by means of $\chi^2$ minimization. The  best estimate for the helium content of blue-MS stars corresponds to the helium abundance of the isochrone that provides the minimum $\chi^{2}$ and corresponds to Y=0.315$\pm$0.010.
%
%
To estimate the uncertainty, we enhanced the color separation between the two MSs by $\pm$1-$\sigma$, where $\sigma$ is the error on the color determination, and derived the corresponding helium values by using the procedure above. The uncertainty corresponds to the average of the absolute differences between these helium determinations and the best helium estimate. \\

The high helium content of the blue MS indicates that it is the counterpart of population 2G$_{\rm B}$, 2G$_{\rm C}$ and 2G$_{\rm D}$ stars. On the contrary, 2G$_{A}$ stars share similar helium content as the 1G (Y$\sim$0.25) and are likely associated with the red MS. 
These conclusions, based on the chemical compositions of MS and RGB stars, are corroborated by the fact that the fraction of 1G$+$2G$_{A}$ stars ($\sim 63$\%) is consistent with the fraction of red-MS stars ($\sim 65$\%).

\subsection{Multiple populations along the Horizontal Branch}
\label{sec:HB}

\begin{figure}
\centering
\includegraphics[width=8.5cm,trim={0.7cm 5.5cm 0.2cm 4.0cm},clip]{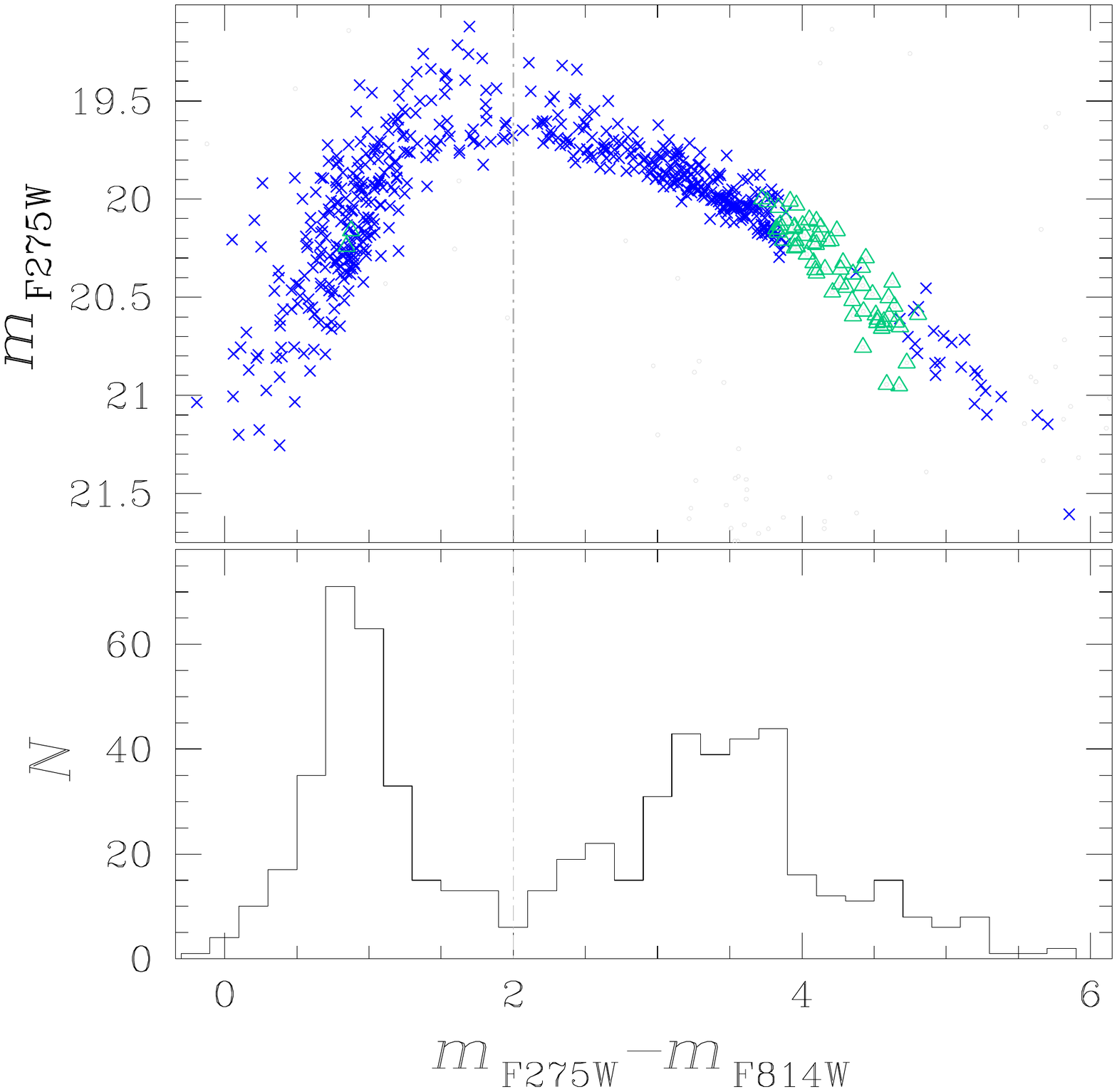}
\caption{$m_{\rm F275W}$ vs.\, $m_{\rm F275W}-m_{\rm F814W}$ CMD of HB stars in NGC\,6402, with RR\,Lyrae marked by green triangles (top). Bottom panels show the histogram color distribution of HB stars. Vertical lines separate the two main groups of HB stars. See text for details.}
\label{fig:HB}
\end{figure}

To investigate multiple populations along the HB, we exploited the $m_{\rm F275W}$ vs.\, $m_{\rm F275W}-m_{\rm F814W}$ CMD of Figure \ref{fig:HB}, which is very sensitive to  the effective temperature and the luminosity of HB stars.
 We identified 629 candidate HB stars, including 65 RR\,Lyrae stars and 27 and 538 stars redder and bluer, respectively, than the  RR\,Lyrae instability strip. The corresponding HB ratio is HBR=0.81\footnote{The HB ratio \citep{lee1994,mackey2005}, is defined as (B$-$R)/(B$+$V$+$R), where B, R and V are the numbers of stars bluer and redder than the instability strip, and the number of variables in the HB, respectively.} and is larger than that derived by \citet[][HBR=0.45]{contreraspena2013}. \\
 
 The histogram distribution of $m_{\rm F275W}-m_{\rm F814W}$ of HB stars  (bottom panel of Figure \ref{fig:HB}) is clearly bimodal with a narrow peak centered at $\rm m_{F275W}-m_{F814W}\sim $0.8 and a much broadened one around $\rm m_{F275W}-m_{F814W}=3.4$.
 Both peaks have colors that differ from those of the \citet{grundahl1999} and the \cite{momany2004} jumps (hereafter G- and M- jumps), which are located around $\rm m_{F275W}-m_{F814W} \sim 2.4$ and $\rm m_{F275W}-m_{F438W}\sim 1.2$. These jumps are universal features of HB morphology and are due to modifications of the stellar atmospheres \citep[e.g.][]{brown2016,brown2017}. 
 The fact that the two peaks  in the HB color distribution appear independent from the G- and M- jumps indicate that they are not related to atmospheric phenomena but can be safely associated to multiple populations. \\
 
 Although the identification of the five populations of NGC\,6402 along the HB is beyond the scope of this paper, we tentatively associate multiple populations along the RGB and the HB as follows.
  We notice that 354 out of 629 HB stars are bluer than $\rm m_{F275W}-m_{F814W}=2.0$. Hence, about the 55$\pm$2\% of HB stars can be tentatively associated with the red HB peak and the remaining 45$\pm$2\% to the blue peak. It is tempting to speculate that the red HB peak is composed of  1G and 2G$_{\rm A}$) while the helium-rich populations (2G$_{\rm B}$, 2G$_{\rm C}$ and, possibly, 2G$_{\rm D}$) evolve into the blue HB peak. Notice anyway that
  these fractions of HB stars are slightly but significantly different from the fractions of RGB stars with pristine helium abundances (62$\pm$2\%) and the fractions of helium rich stars (38$\pm$2\%).
 
\begin{table*}
    \centering
    \caption{Properties of 1G and 2G$_{\rm E}$ stars along the HB of NGC\,6402. Columns are: population ID, helium abundance (Y), average mass loss ($\mu$), mass loss spread ($\delta$), stellar mass at the tip of the RGB ($\rm M_{\rm Tip}$) and average HB mass ($\rm \bar{M}_{HB}$).  }
    \begin{tabular}{cccccc}
        \hline
        \hline
        ID & Y & $\rm \mu/M_\odot$&$\rm \delta/M_\odot$&$\rm M_{Tip}/M_\odot$&$\rm \bar{M}_{HB}/M_\odot$\\
        \hline
        \hline
        1G  & 0.250 &$0.240\pm 0.022$&$0.006\pm 0.002$& 0.841&$0.601\pm 0.022$\\  
        2G$_{\rm C}$ & 0.315 &$0.280\pm 0.024$&$0.005\pm 0.002$& 0.757&$0.481\pm 0.024$\\ 
        \hline
        \hline
    \end{tabular}
    \label{tab:par_hb}
\end{table*}

\subsubsection{Mass loss of multiple populations}
\label{sub:extremes}
\begin{figure}
    \centering
    \includegraphics[width=\columnwidth,trim={0cm 1cm 0cm 0cm}]{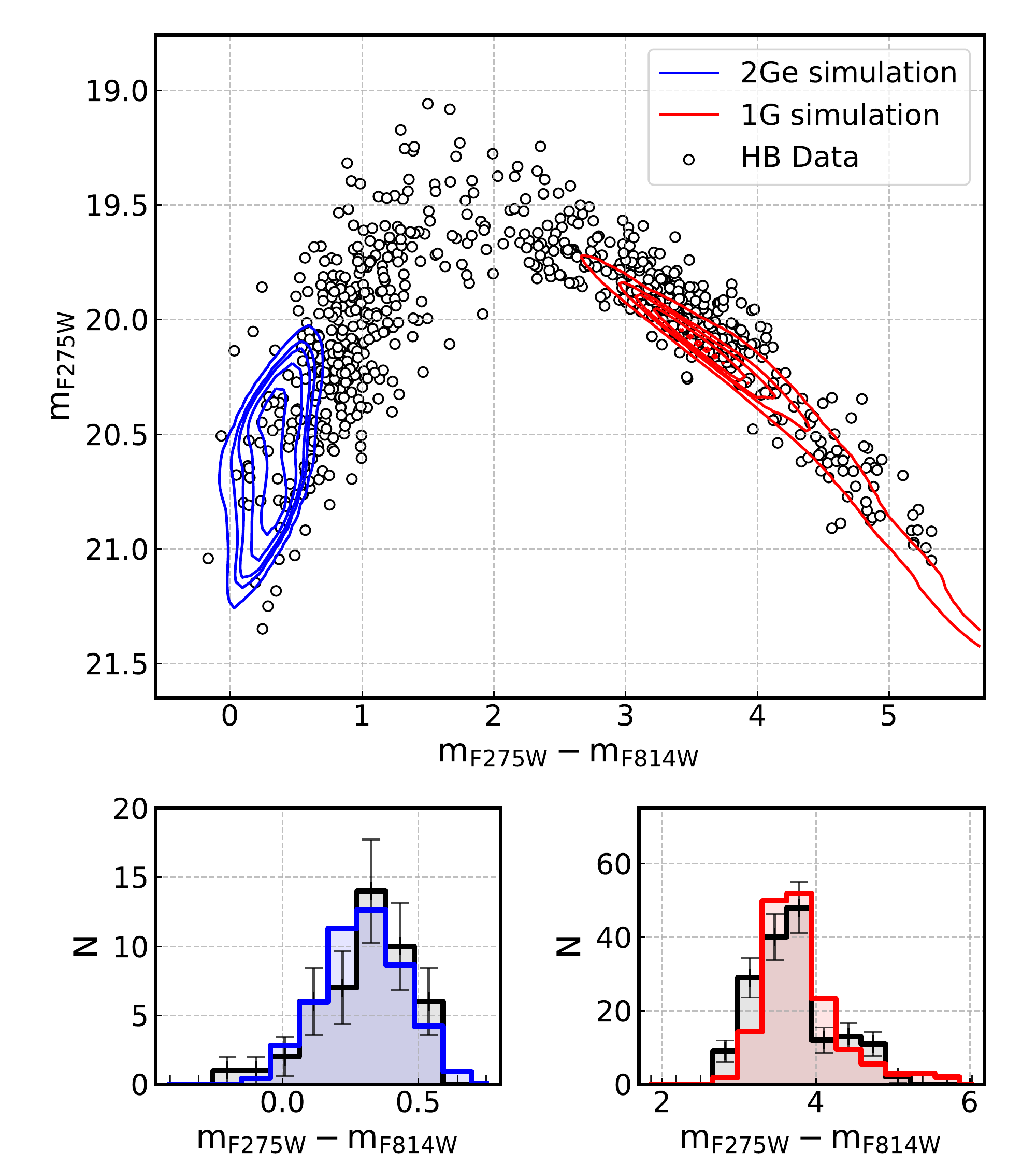}   
    \vskip 10pt 
    \caption{\textit{Upper panel}: $\rm m_{F275W}-m_{F814W}$ vs. $\rm m_{F275W}$ CMD of the HB stars in NGC\,6402. The contour plots represent the best fit simulation of the 1G and the extreme 2G, respectively for red and blue. The two bottom panels represent the histograms of the colour distributions of observed (black) and simulated stars as direct comparison with the part of the HB the two simulations overlap in the CMD. }
    \label{pic:extremes_6402}
\end{figure}

Work based on both theory and high-resolution spectroscopy reveals that the reddest HB tail is composed of the bulk 1G stars, while the bluest portion of the HB is populated by the 2G stars with extreme helium content \citep[e.g.][]{dantona2002, marino2011m4}.
Based on this evidence \citep[][]{tailo2019, tailo2020, tailo2021} identified 1G stars and 2G stars with extreme chemical composition in 56 Galactic GCs and measured the RGB mass loss.
 In the following, we extend to NGC\,6402 the method introduced by Tailo and collaborators to identify 1G and 2G$_{\rm C}$ stars along the HB and infer their RGB mass losses. \\
 
 In a nutshell, the method consists of comparing the observed color distributions of the reddest and bluest HB stars with the colors of a grid of synthetic HB stars. Simulated HBs have different average mass loss ($\mu$) and mass loss spread ($\delta$) values. Specifically, $\mu$ ranges from 0.200 $\rm M_\odot$ to 0.310 $\rm M_\odot$ in steps of 0.003 $\rm M_\odot$ and  $\delta$\ varies from 0.002 $\rm M_\odot$ to 0.012 $\rm M_\odot$ in steps of 0.001 $\rm M_\odot$.
 
 We assumed for all simulations [Fe/H]=$-$1.1, [$\alpha$/Fe]=0.3 and age of 12.5 Gyr. We adopted Y=0.25 for 1G stars and Y=0.315 for the 2G$_{\rm C}$. Further details on the procedure are provided by \citet[][]{tailo2020, tailo2021}. \\

The simulation of 1G stars that provides the best match with the histogram distribution of the reddest HB stars is derived by means of $\rm \chi^2_d$ minimization and corresponds to $\rm \mu=0.240\pm 0.022\, M_\odot$ and $\rm \delta=0.006\pm0.002\, M_\odot$. Similarly, we obtain for 2G$_{\rm C}$ stars $\rm \mu=0.280\pm0.024\, M_\odot$ and $\rm \delta=0.005\pm0.002 M_\odot$. 
 Results are listed in Table \ref{tab:par_hb} and plotted in Figure \ref{pic:extremes_6402}, where we superimposed the contours of the best-fit synthetic CMDs on the observed CMD (upper panels) and compare the observed and simulated histograms of the colors of 1G and 2G$_{\rm C}$ HB stars. \\
 
\begin{figure}
    \centering
    \includegraphics[width=\columnwidth,trim={0cm 1cm 0cm 0cm}]{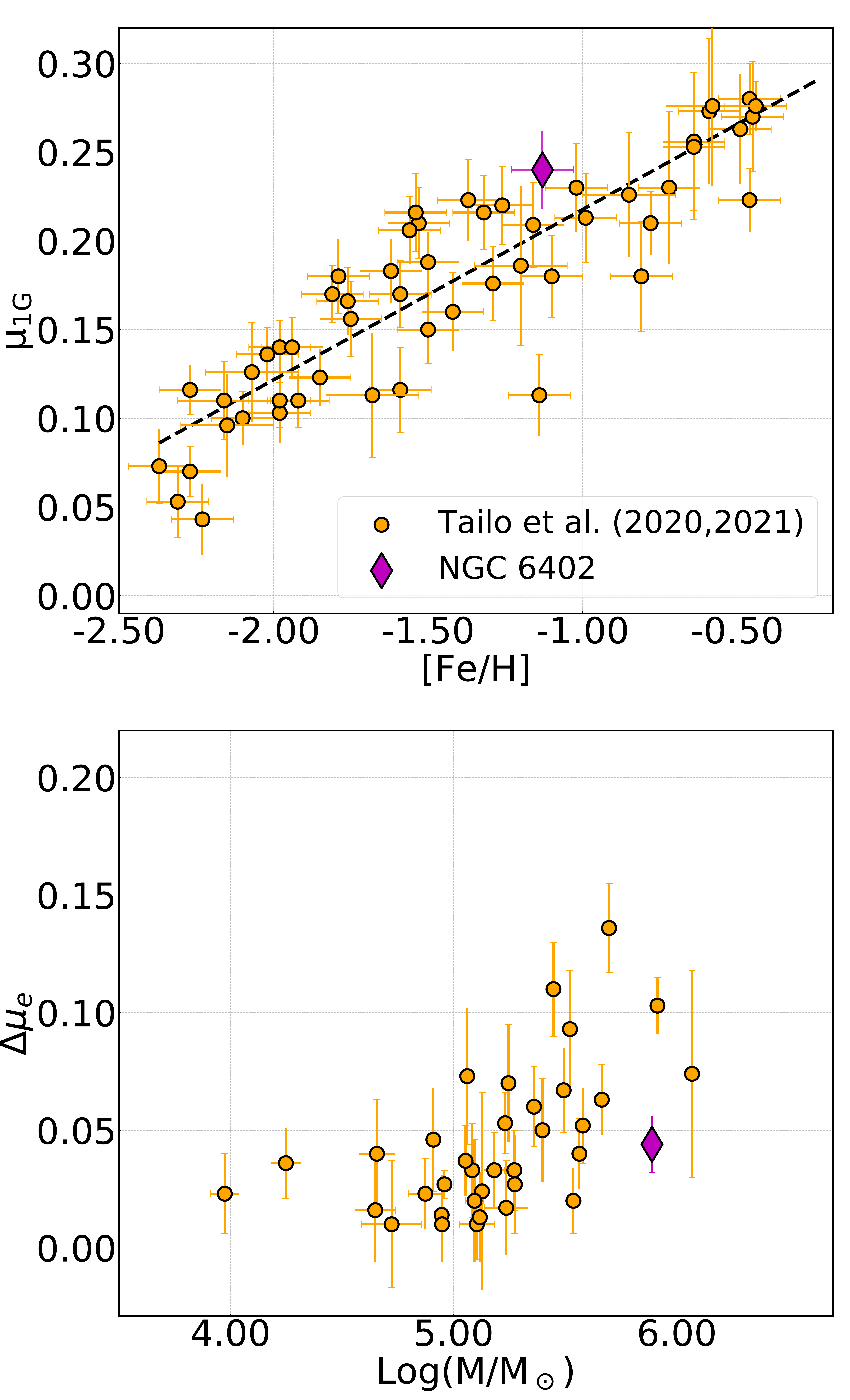}  
    \caption{
    Diagrams of the mass lost during the RGB evolution, $\rm \mu_{1G}$,  versus [Fe/H] (top) and  of the extra--mass loss of the 2G, $\rm \Delta \mu_e$, versus the cluster mass Log($\rm M/M_\odot$)  (bottom), from \citet[][]{tailo2020, tailo2021}. $\rm \mu_{1G}$\ and 
$\rm \Delta \mu_e$\ are i solar masses.  NGC\,6402 is marked with purple diamonds. The black dashed line in the top panel is the best fit least squares straight line of all points. }     
    \label{pic:mu1g_6402}
\end{figure}

Interestingly, the amount of mass lost by 1G stars in NGC\,6402 is comparable with the mass loss of 1G stars in clusters with similar metallicity. Indeed, as shown in the top panel of Figure \ref{pic:mu1g_6402}, NGC\,6402 follows the same trend in the [Fe/H] vs. $\rm \mu_{1G}$ plane discovered by \citet{tailo2020,tailo2021}. 
We also find that 2G$_{\rm C}$ lose more mass than the 1G, in analogy with what is observed in nearly all massive GCs. This result is shown in the bottom panel of Figure \ref{pic:mu1g_6402}, where NGC\,6402 is represented by the purple diamond, and the dots show the extra mass loss $\rm \Delta \mu_e$\ in the sample of GCs studied by \citet{tailo2020}.
%
%
%
Actually, the bottom panel of Fig.\,\ref{pic:mu1g_6402} shows that the extra mass loss ($\Delta\mu_e=0.040\pm 0.012\, M/M_\odot$) for NGC\,6402 is relatively small with respect to that of the other massive clusters. 
The result depends on the helium content Y=0.315 assumed for 2G$_{\rm C}$\ stars: for a smaller value of Y, $\Delta\mu_e$\  would be larger and fit better with the trend defined by bulk of the other clusters. We take this as an indication that Y=0.315 may represent an upper limit to the possible helium of this extreme population.
\begin{table*}
  \caption{Fractions of populations  in different samples}
\center
\begin{tabular}{l l c c  c c c c c c}
\hline \hline
Name &\multicolumn{2}{c}{J2019} 
&\multicolumn{2}{c}{~~~HST Parallel field~~~}  &  \multicolumn{2}{l}{~~HST ChM~~} & Tot ChM &\multicolumn{2}{c}{HB}  \\
\hline
1G &P1      &   0.34     &   &&                 1G &  0.280  \\
       &&& ~~red MS    & 0.652      &      & &0.626 &red   &  0.55  \\ 
2G mild & P2    &     0.40   &                      &  &        2G$_{\rm A}$       &0.346  && ($m_{\rm F275W}-m_{\rm F814W}>2$) &  
\\
\hline
2G extreme & E    &     0.26   &      ~~blue MS  & 0.342                  &       2G$_{\rm B}$ &   0.161  &0.374 &blue  &  0.45 \\
&&&&&   2G$_{\rm C}$  &  0.129 && ($m_{\rm F275W}-m_{\rm F814W}<2$) &\\
&&&&&  2G$_{\rm D}$   &  0.084 \\
     \hline\hline
\end{tabular}
  \label{tab:ratios}
 \end{table*}

\section{Summary on the fraction of different populations from different data}

Table\,\ref{tab:ratios} compares the information gathered on the different populations. The results from the spectroscopic sample in \cite{johnson2019} are given in the second column. The HST parallel field data include 65\% stars in the red MS and 35\% stars in the blue MS. The red MS includes both the P1 and P2  populations. The comparison with the ChM data is quite consistent, if we add together the 1G and 2G$_{\rm A}$ data (see Fig.\,\ref{fig:chimica}) and collect as a generic extreme E group all the other populations, including the 2G$_{\rm D}$ with probably higher metallicity. In general, we may say that the E group is larger than found in \cite{johnson2019}, a result not so surprising when accounting for the fact that the spectroscopic sample is quite small. The HB data are subdivided into ``red" and ``blue", at the right and left side of the color m$_{\rm F275}$--m$_{\rm F814}=$2  (Fig.\,\ref{fig:HB}). The resulting fractions are perhaps a bit more surprising, because here we have an even larger --45\%-- fraction of stars belonging to the blue side of the HB, which should be assigned to the E populations. A possible explanation of the higher percentage of blue HB stars may be that our fraction of red HB stars is too small, as testified by the fact that our RR\,Lyr group is much less represented than in \cite{contreraspena2018}. At first sight, we could attribute the different fraction to the fact that our RR\,Lyr sample is obtained at the center of the cluster, while the RR\,Lyrae in the whole catalogue are spread all over the cluster: it is plausible that the 1G stars are less concentrated than the 2G stars, as we know to occur in several GCs
 \citep[see e.g.][]{sollima2007, bellini2009, lardo2011, milone47tuc2012, simioni2016, dalessandro2019, dondoglio2021};
 \cite{dalessandro2019}  find a spatial segregation of the 2G in about half of a sample of 20 clusters, correlated with the current half mass relaxation time $t/t_{rh}$. NGC\,6402, with an age $t$=12.5\,Gyr derived here, and $t_{rh}\sim 2.5\times 10^9$yr \citep{harris1996} has $t/t_{rh} \sim 5$, in the range where segregation should still be present, and deserves a study of the spatial distribution of the two populations. Anyway, the parallel field is 6' far from the cluster center, well out of the central region, but it shows a proportion of 1G+2G mild and 2G extreme stars consistent with those derived from the core data from which the ChM is derived. Thus, further investigation is needed to understand the meaning of the results concerning the HB. We conclude that the HST sample contains about 35\% of stars belonging to the extreme and probably also intermediate populations, while another $\sim$35\% of stars belong to the mild 2G.

\begin{figure*}
\centering
\includegraphics[height=5.4cm,trim={0.75cm 5.cm 7.5cm 12.0cm},clip]{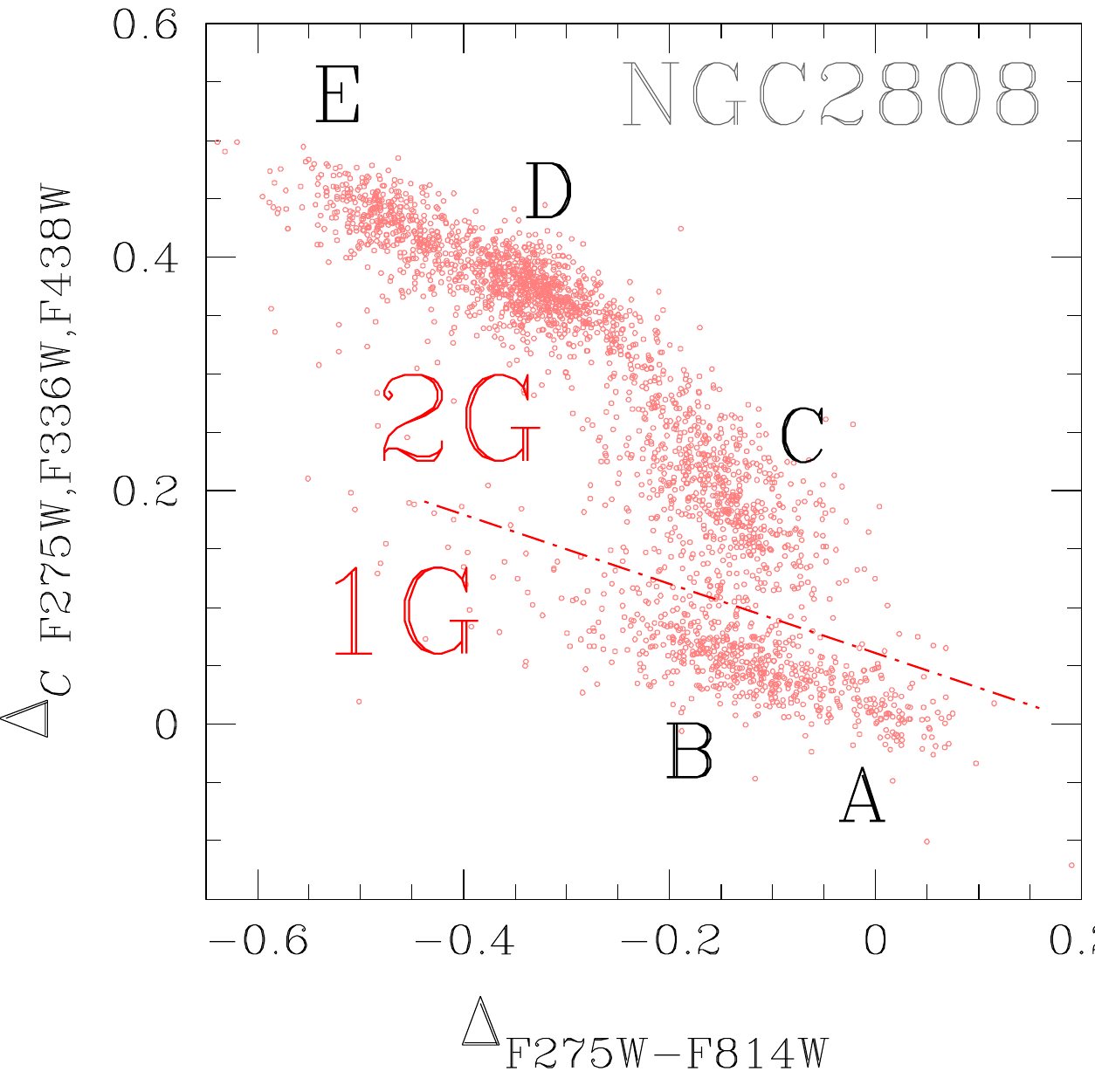}
\includegraphics[height=5.4cm,trim={0.7cm 5cm 3.0cm 12cm},clip]{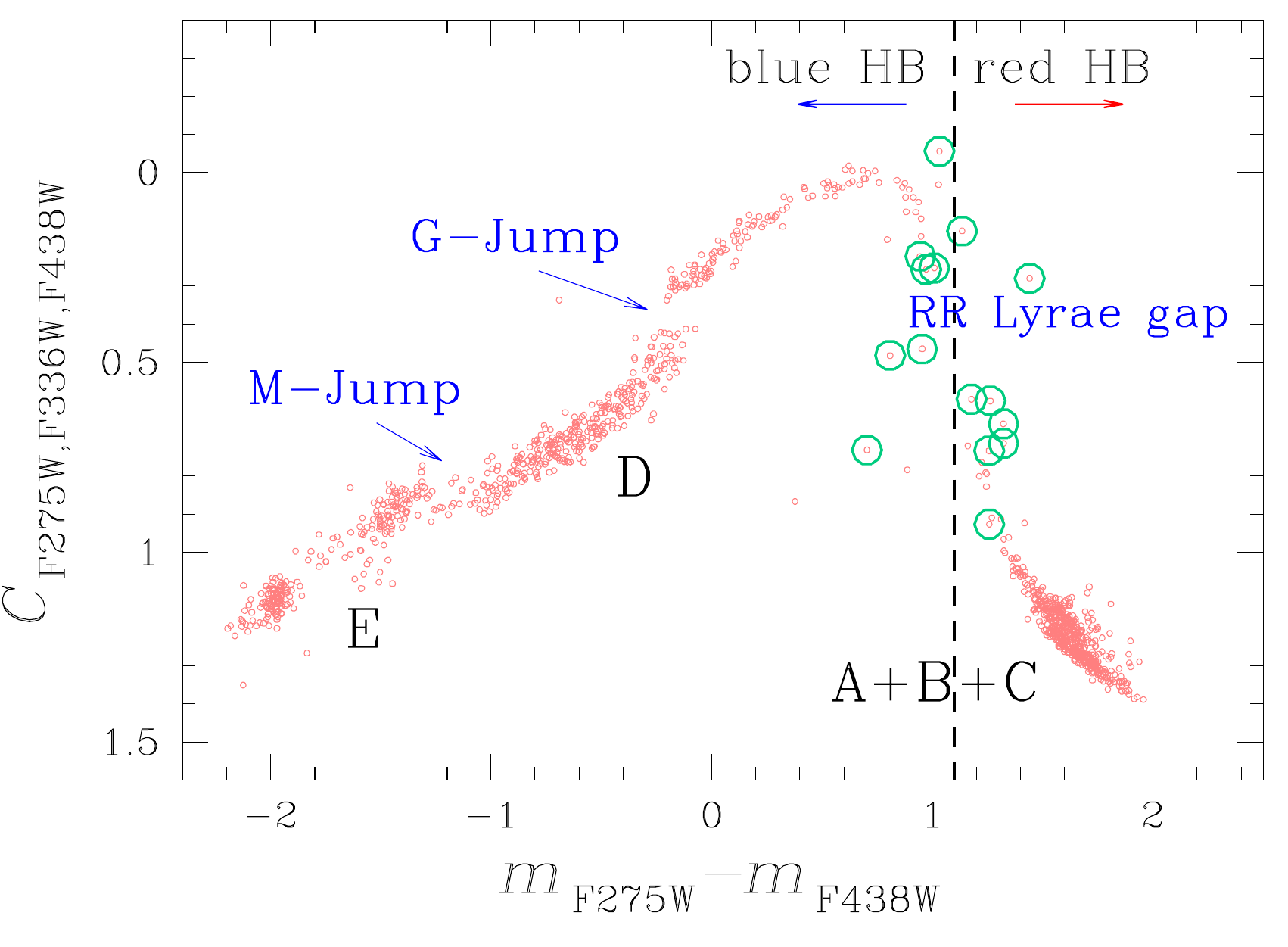}
\includegraphics[height=5.45cm,trim={0.8cm 5cm 7.5cm 10cm},clip]{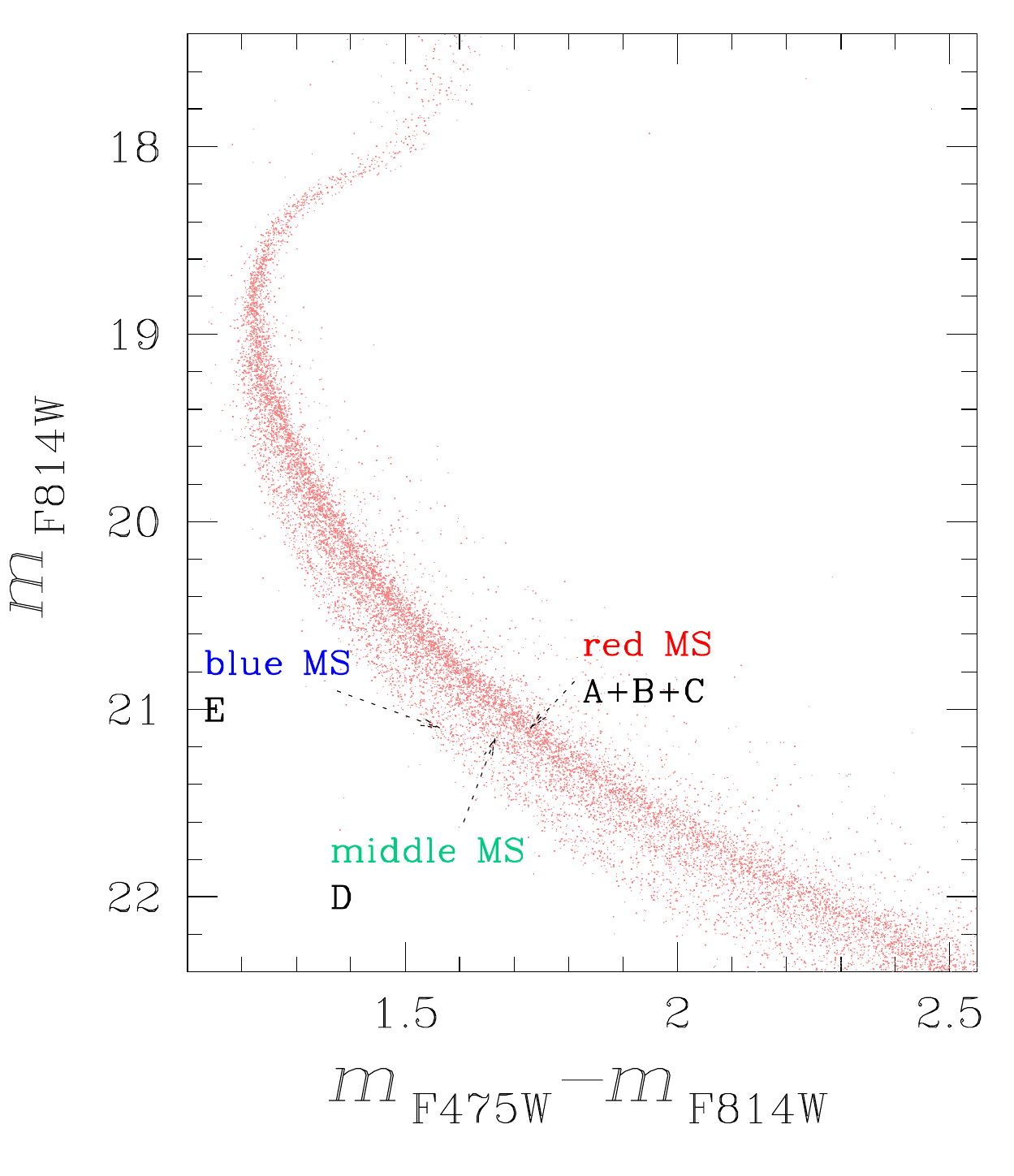}
\caption{\textit{Left}. ChM of RGB stars in NGC\,2808. The red dashed-dot line separates the bulk of 1G stars from the 2G.  \textit{Middle}. $C_{\rm F275W,F336W,F438W}$  vs.\,$m_{\rm F275W}-m_{\rm F814W}$ diagram of HB stars in NGC\,2808. Candidate RR\,Lyrae stars are marked with green circles. The vertical dashed line separates the red and the blue HB. We indicated the position of the M- and G- jumps and the gap associated with the RR\,Lyrae instability strip.
\textit{Right}. Optical, $m_{\rm F814W}$ vs.\,$m_{\rm F475W}-m_{\rm F814W}$, CMD of NGC\,2808 stars zoomed on the MS and the SGB. The red, middle and blue MS are indicated by the black arrows.
The letters A--E in all panels indicate the five main populations of NGC\,2808.
}
\label{fig:n2808}
\end{figure*} 
\section{Comparison with NGC\,2808}
\label{sec:2808comp}

In this section we directly compare the photometric diagrams of NGC\,6402 and NGC\,2808, which is one of the most-studied  GCs in the context of stellar populations. Such comparison is  possible because NGC\,2808 and NGC\,6402 share similar metallicities, ages, and masses. \\ 

NGC\,2808 hosts populations with extreme contents of helium and light elements that have been investigated along various evolutionary sequences, including the MS, RGB, HB, and even the AGB \citep[e.g.][]{dantona2005, piotto2007, carretta2009a,  marino2014hb2808, marino2017a, lagioia2021a}.  In the following, we discuss multiple populations in NGC\,2808 along the RGB stars, the MS, and the HB by using the same photometric diagrams available for NGC\,6402. \\

As illustrated in the left panel of Figure \ref{fig:n2808}, the ChM of NGC\,2808 reveals that both 1G and 2G stars are not chemically homogeneous. 1G stars host two main stellar populations (namely A and B), whereas the 2G is composed of at least three distinct groups of stars (C, D and E).
 1G stars share pristine helium abundance (Y$\sim$0.25) and light-element abundances but are possibly inhomogeneous in metallicity, with population-A stars being slightly more metal rich than population-B stars.
 On the contrary, 2G stars are enhanced in He, N, and Na and depleted in C and O. 
  Specifically, population-C stars may be very slightly enriched, by $\sim$0.005\,dex in helium mass fractions \citep{dantona2016}, while population-D and E stars have larger helium contents up to Y$\sim$0.29, and Y$\sim$0.35, respectively. \\

NGC\,2808 exhibits an extended HB that is well populated on both sides of the RR\,Lyrae instability strip. The $C_{\rm F275W,F336W,F438W}$  vs.\,$m_{\rm F275W}-m_{\rm F814W}$ diagram of HB stars is plotted in the middle panel of Figure \ref{fig:n2808}, where we marked the position of the G- and M-jumps.
Work based on high-resolution spectroscopy and on simulated HBs show that the red HB host stars with different sodium contents but similar helium abundances. Hence, it is composed of population A, B, and C stars \citep[e.g.][]{ marino2014hb2808, dantona2016}.  Population-E stars mostly evolve into the bluest HB tail, on the hot side on the M-jump. The HB region between the RR\,Lyrae and the M-jump is mainly composed of population-D stars. \\

CMDs made with optical filters  do not allow us to disentangle the five stellar populations of NGC\,2808.  
Nevertheless, the $m_{\rm F814W}$ vs.\,$m_{\rm F475W}-m_{\rm F814W}$ CMD plotted in the right panel of Figure \ref{fig:n2808} reveals a triple MS. In this diagram, the red MS is composed of stars of the populations A, B, and C and is reproduced by isochrones with nearly pristine helium content.  The middle and the blue MS correspond to population-D and population-E stars, respectively. 
From \citet{milone2012mf2808}, the three MS populations contain the following fractions of stars:  0.62$\pm$0.02 (red MS), 0.24$\pm$0.02 (middle MS) and 0.14$\pm$0.03 (blue-MS).

\begin{figure*}
\centering
\includegraphics[height=6.7cm,trim={0.7cm 5.3cm 6.7cm 12cm},clip]{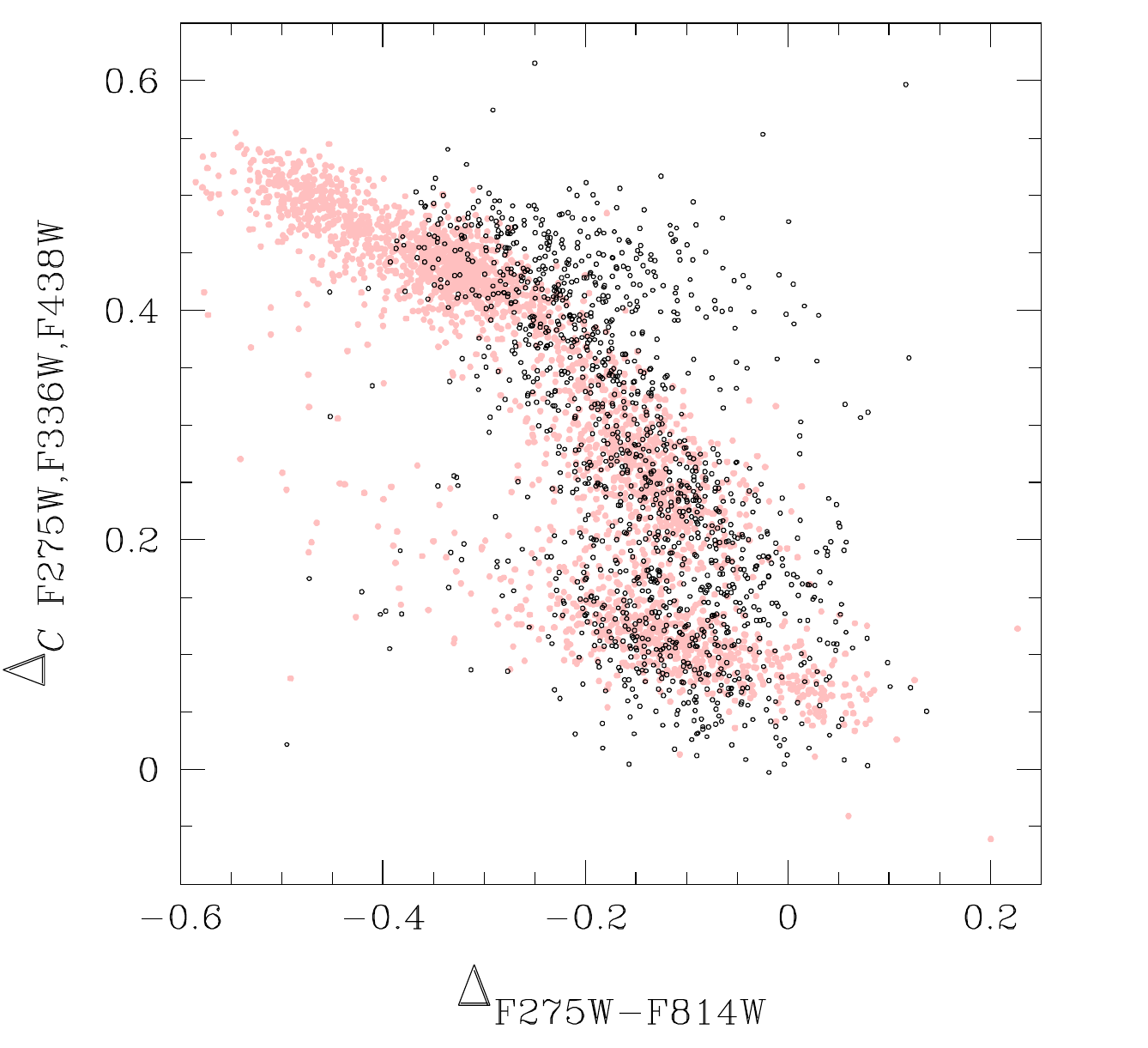}
\includegraphics[height=6.6cm,trim={0.0cm 4.5cm 0.00cm 9.1cm},clip]{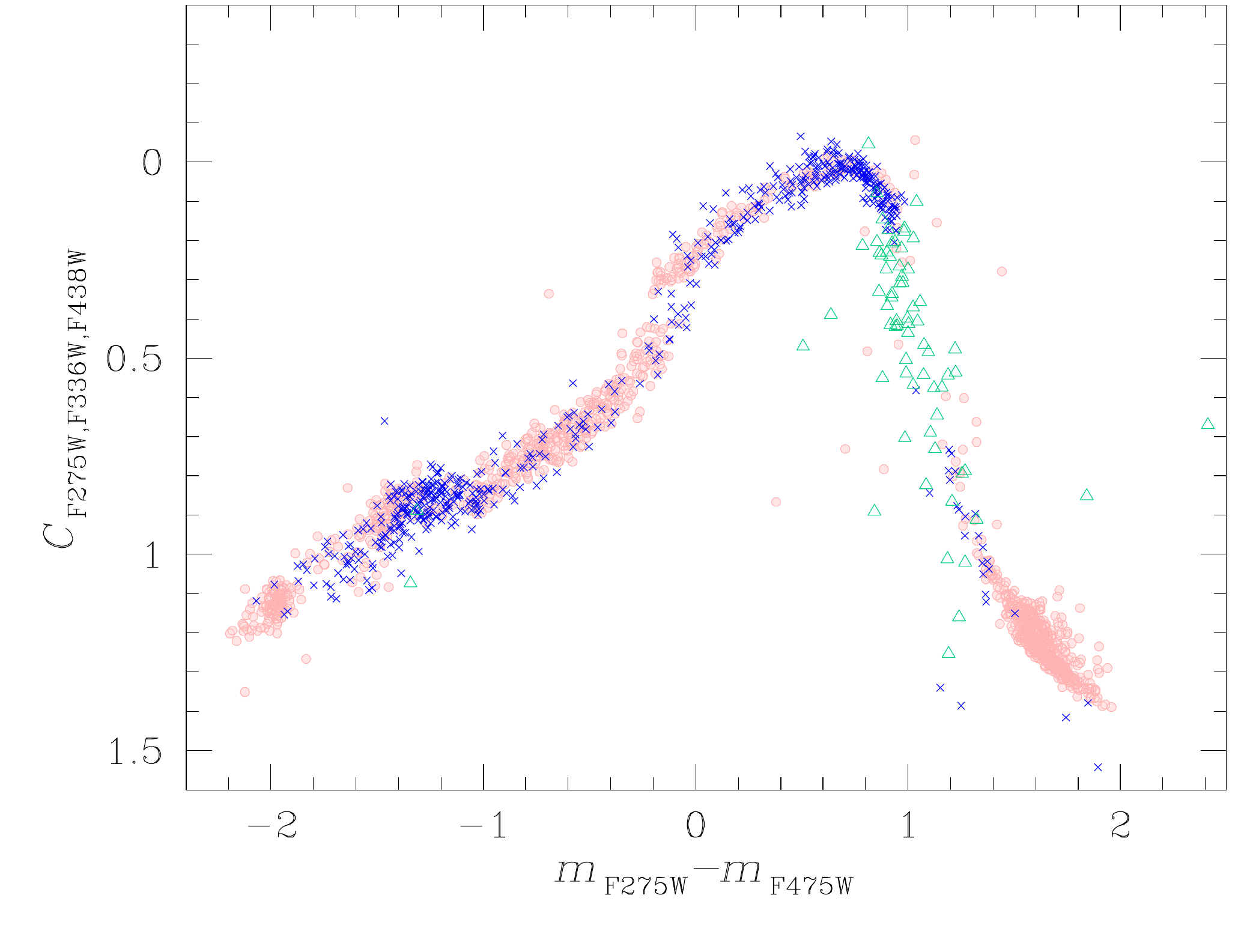}
\includegraphics[height=6.7cm,trim={0.7cm 5.3cm 7.5cm 12.0cm},clip]{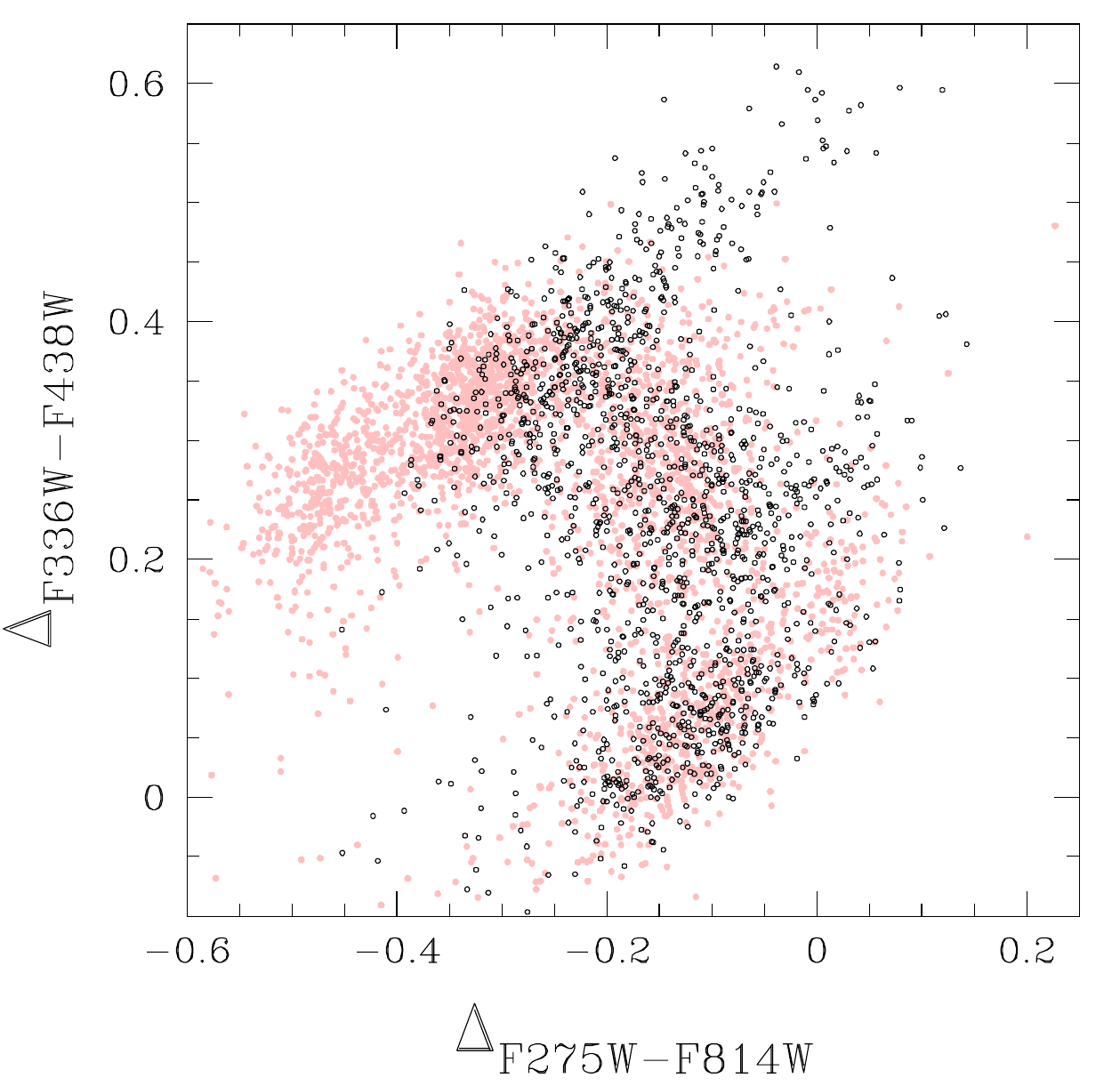}
\includegraphics[height=6.7cm,trim={0.7cm 5.5cm 0.0cm 10.2cm},clip]{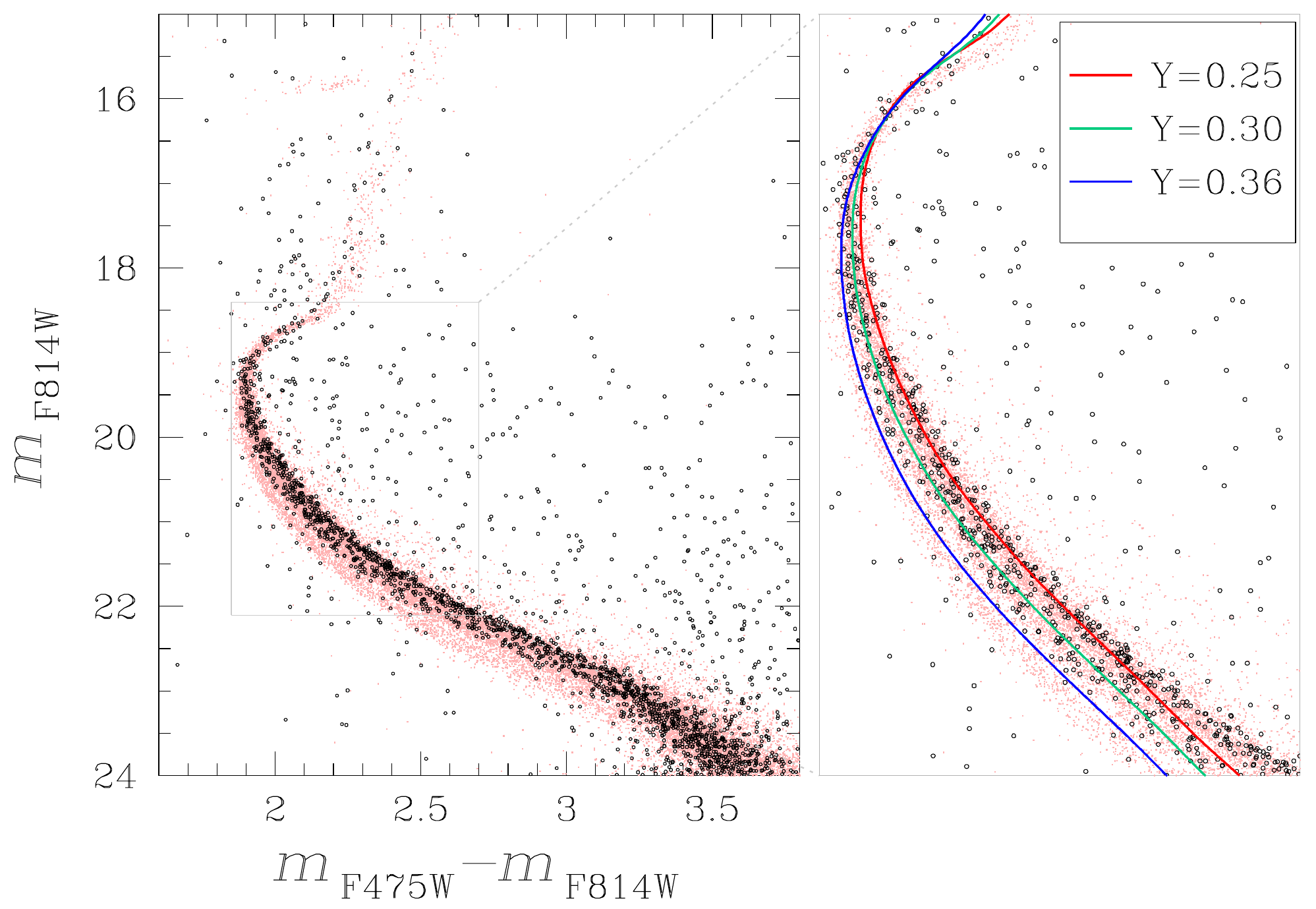}
\caption{\textit{Upper panels.} Comparison of the $\Delta_{C, {\rm F275W,F336W,F438W}}$ vs.\,$\Delta_{\rm F275W-F814W}$ (upper-left) and  $\Delta_{\rm F336W-F438W}$ vs.\,$\Delta_{\rm F275W-F814W}$ ChMs (lower-left) of NGC\,2808 (pink points) and NGC\,6402 (black points). Right panels compares the HBs of NGC\,2808 (pink) and NGC\,6402 (same colors as in Figure \ref{fig:cmd}, top)  and the $m_{\rm F814W}$ vs.\,$m_{\rm F475W}-m_{\rm F814W}$ CMDs (bottom). We also provide a zoom of the CMD on the upper MS, where we compare the observations of NGC\,2808 and NGC\,6402 with isochrone with different helium abundances. See text for details.  }
\label{fig:Comp2808chm}
\end{figure*}

The comparison between NGC\,2808 and NGC\,6402 is illustrated in Figure \ref{fig:Comp2808chm}. In the upper-left panel we superimposed their $\Delta_{C {\rm F275W,F336W,F438W}}$ vs.\,$\Delta_{\rm F275W-F814W}$ ChMs in such a way that 1G stars of both clusters share the same median values. \\

Clearly, the ChM of NGC\,6402 is significantly less extended in both axis with respect to the NGC\,2808 map. In particular, there is no evidence for NGC\,6402 counterparts of the most helium-rich stars of NGC\,2808 (population E).  
We also find that 1G stars of NGC\,6402 roughly match the population B of NGC\,2808, while there is no evidence for population-A like stars.
The group of 2G$_{\rm A}$ stars of NGC\,6402 mostly overlaps the population C of NGC\,2808 and the similarity between these populations is corroborated by their chemical compositions. Indeed, both 2G$_{\rm A}$ stars of NGC\,6402 and population-C stars of NGC\,2808 share similar He and O content as the 1G but are significantly enhanced in nitrogen.
Populations  2G$_{\rm B}$ and 2G$_{\rm C}$  are analogs of population-D stars in NGC\,2808, as demonstrated by the location on the ChM and the helium content $Y \sim 0.30$. 
Intriguingly, NGC\,2808 lacks a counterpart of the 2G$_{\rm D}$ stars. These results are confirmed from the comparison of the $\Delta_{\rm F336W-F438W}$ vs.\,$\Delta_{\rm F275W-F814W}$ ChMs of NGC\,2808 and NGC\,6402 plotted in the bottom-left panel of Figure\,\ref{fig:Comp2808chm}. \\

The lack of stars with extreme helium content in NGC\,6402 is corroborated by the comparison of the $m_{\rm F814W}$ vs.\,$m_{\rm F475W}-m_{\rm F814W}$ CMDs (bottom-left panel). Clearly, the red and middle MS of NGC\,2808 match the red and blue MS of NGC\,6402, whereas NGC\,6402 does not host the counterpart of the blue MS (i.e. the population E) in NGC\,2808. As expected, when we superimposed on the CMD the isochrones with different helium contents that reproduce the triple MS of NGC\,2808, it results that blue-MS stars of NGC\,6402 are not consistent with the most helium rich isochrone (Y=0.36). \\

The comparison of the HBs of the two clusters is plotted in the upper-right panel of Figure\,\ref{fig:Comp2808chm}. Clearly, HB stars of both clusters follow the same sequence in the $C_{\rm F275W,F336W,F438W}$  vs.\,$m_{\rm F275W}-m_{\rm F814W}$ plane \citep[see][for details]{brown2016}.
The HB of NGC\,6402 exhibits a shorter $m_{\rm F275W}-m_{\rm F814W}$ color extension compared to NGC\,2808 and shows no evidence for blue-hook stars, which are the progeny of stars highly enhanced in helium. These facts corroborate the evidence that NGC\,6402 does not host stars with extreme helium content.
At odds with NGC\,2808, NGC\,6402 does not show the red HB, thus NGC\,6402 is a `second parameter' cluster belonging to the M\,13 group. For a recent discussion of this problem see \citet{tailo2020} and references therein. Since the reddest HB tail is mostly populated by 1G stars, we expect, in addition to metallicity, at least one second parameter is responsible for the color of 1G stars. Age difference as well as mass loss differences are possible second parameters. As an alternative,   stars on the red HB tail of NGC\,6402 should be more helium rich than the red-HB stars of NGC\,2808 \citep[][]{dc2008}.    

\section{A summary of models}

The issue of multiple populations includes chemical `anomalies' of quite a number of elements, with different problems highlighted by different stars in different clusters. For some of these anomalies the contribution of some particular polluters may be required\footnote{For instance, the s-process elemental distribution in  the metal poor star ROA\,276 in \ocen \citep{yong2017} is well explained by the rotating models by  \cite{frisch2012, frisch2016}. }.
Nevertheless, most of the scenarios proposed to explain the light element variations do not withstand a general scrutiny.  
We refer the reader to the complete summaries available in the recent literature \citep[e.g.,][]{renzini2015,bastian_lardo2018, gratton2019} and discuss here the only two remaining models which at least basically conform to the requirements of the chemistry displayed in massive clusters, such as NGC\,2808 and NGC\,6402, containing an ``extreme" population. These are the super-massive star (SMS) model, here considered in the basic formulation by \citet{denissenkov2014} and \citet{denissenkov2015} that also includes a qualitative description in terms of a dynamical scenario \citep{gieles2018}, and the AGB model \citep{ventura2001}. In the AGB stars the hot environment where the light elements are processed by proton captures is not the stellar core, but burning at the ``hot bottom" of the convective envelope (HBB), from where the elements are then transported by convection in the whole envelope and then lost by stellar wind and planetary nebula ejection.\\ 
For the present work, the models are considered mainly to understand whether the ejecta composition  (with or without dilution with pristine gas) is compatible both with the patterns of light element abundances in the giants \citep{johnson2019} and with the helium content  derived here for NGC\,6402. 
At the end of our analysis, we will conclude that neither model in its present stage  of  development  is  compatible  with  the  helium abundance  of  the  extreme  2G,  and  suggest possible additional formation mechanisms.

\subsection{The SMS model}
\label{subsec:SMS}
The reason to propose that ``supermassive'' stars of about 10$^4$\msun\ are at the origin of the elemental abundances found in the second population of GCs, although the highest mass which can be formed by simple fragmentation is 100 times smaller, is that such structures achieve the central temperature of $\sim$75\,MK, necessary to allow the reaction $^{24}$Mg(p,$\gamma$)$^{25}$Al which depleted $^{24}$Mg in the gas forming the extreme GC stars. This temperature is not reached in the cores of standard massive stars. Another advantage is that the p-captures by Mg isotopes in the SMS cores produce isotopic ratios compatible with the ratios observed in five clusters \citep{dacosta2013}, while discrepant ratios are found when these reactions occur in the AGB HBB environment \citep{ventura2009}\footnote{The correct isotopic ratios for magnesium can be obtained in the AGB models by enhancing by a factor $\sim$3 the proton-capture rate by $^{25}$Mg at the temperatures
$\sim$100\,MK, in particular the $^{25}$Mg(p, $\gamma$)$^{26}$Al$^{m}$ channel, beyond the most recent experimental determinations \citep{ventura2018mgisotopes}. With this input, the AGB nucleosynthesis also correctly predicts the total Mg depletion in different clusters, and {\it its trend with the metallicity}. Larger Mg depletions are observed in clusters of smaller metallicity (e.g. Mg is reduced by a factor $\sim$10 in cluster NGC\,2419 \citep{cohenkirby2012}). This trend is an observational constrain not directly predictable in the case of SMS, as the temperature at which nuclear activity takes place in the central regions is almost unaffected by the metallicity, unlike the HBB conditions in massive AGB stars.}. Difficulties remain to explain the production of silicon by p--captures on aluminum ---a Si-Mg anticorrelation is present in NGC\,2808 giants \citep{carretta2015}--- as Si production occurs at the expense of the total depletion of Mg and Na in the cores of SMS; further,  potassium can not be produced \citep{prantzos2017}.
Nevertheless, the SMS model also provides an advantage in the description of the oxygen depletion and sodium enhancements (see in Sect.\,\ref{sub:agbmodel} the difficulties of the AGB model) as shown in the case of M\,13 by  \cite{denissenkov2015}, whose 60$\times 10^3$\msun\ model chemistry at Y$_{\rm core}$=0.38, diluted by 50\% with pristine gas is able to reproduce the abundance patterns O--Na and the isotopic Mg ratios.

Notice that for these same O--Na abundances, a small dilution by $\sim$10\% with pristine matter brings the O-Na abundances in the correct range to describe the E stars in NGC\,6402. \\

A further interesting feature of the SMS model is that it could help solve the ``uniqueness" problem of extreme composition stars being found only in a few very massive old GC, because only those had the opportunity to host and evolve SMS (see in Sect.\,\ref{sub:agbmodel} the alternative view of the AGB model).  \\ 

Going beyond the nucleosynthesis results, we must remark that the possible successes of the SMS site for nucleosynthesis of multiple populations has strong constraints, namely:
\begin{enumerate}
    \item The protocluster must be able to form SMS of about the required mass of $\sim$2--10$\times 10^3$\msun\ \citep{denissenkov2014, prantzos2017}. First, massive stars must sink to the cluster core due to dynamical friction and then merge by multiple collisions. Models aimed at explaining the possible presence of intermediate mass black holes in young massive clusters have been developed  \citep{portegies1999, portegies2004}. The process is constrained to occur within the very short main sequence timescale of such objects. 
    \item The core H-burning (main sequence)  must be stopped very early, and  at the specific time required to avoid that the SMS ejecta have helium content larger than the maximum helium observed in the extreme GC populations, roughly constrained to be always Y$<$0.4. The hypothesis  is  that the SMS lose at early times a great fraction of their mass, as a result of some instabilities and stellar winds. No quantitative models are still available to support these suggestions.
    The ejecta then mix with the gas of the environment and form the second population stars, showing different degrees of p-capture processing.
\end{enumerate}
For the purpose of this work, we put aside these problems, and keep the chemistry of the diluted SMS gas as the basis for the formation of the E population in NGC\,6402, as depicted in \cite{johnson2019}. Note anyway that the E population includes about 26\% of the cluster mass, $\sim 3.5 \times 10^5$\msun\ if we refer to a present mass of 1.4$\times 10^6$\msun. Assuming now a dilution of roughly $\sim$40\% with pristine gas, we need anyway to process within SMS $\sim 2\times 10^5$\msun. Thus, either 20 SMS of 10$^4$\msun\ each, or a model such as the ``conveyor belt" proposed by \cite{gieles2018} is necessary to achieve such a result. We must therefore consider that there are still many unexplored aspects in the model.

\subsection{The AGB model}
\label{sub:agbmodel}
The AGB model attributes the chemistry of multiple generations to star formation in the gas expelled at low velocities by massive AGB and super--AGB stars, collected into the central regions of the cluster and possibly diluted with residual or re-accreted pristine gas.  \\

From the point of view of chemistry of the different multiple populations, \cite{dantona2016} have developed the idea that they are the result of formation along the ``timeline" of evolution of the different super-AGB and AGB masses, having different ejecta composition. Application to the prototype cluster NGC\,2808 showed that the populations A, B, C, D, E by \cite{milone2015} should have formed in the order B E D C A. At first, the standard population B forms, then the extreme population E, {\it undiluted}, at the time of evolution of the masses which undergo super--AGB evolution, then the intermediate population D, partially diluted with pristine gas, then the population C, very diluted with pristine gas and thus showing only very mild differences with respect to the standard B. The last population A was suggested to be polluted by the ejecta of some Type\,Ia supernovae, whose regular explosions eventually expel the re-accreting pristine gas and end the second star formation epoch. Of course the details of the evolution of each cluster determine the variety of chemical patterns found among different clusters. In particular, the E population, whose chemistry indicates that it forms {\it before} re-accretion, can be present only in clusters in which the re--accretion of pristine gas is delayed enough, and models show that these are the most massive clusters only \citep{dercole2016}, in agreement with the observations \citep[e.g. Fig.\,13 in][]{milone2018he}.
\begin{figure}
\centering
\includegraphics[width=8.5cm, trim={0.7cm 6.5cm 1.5cm 4.5cm},clip]{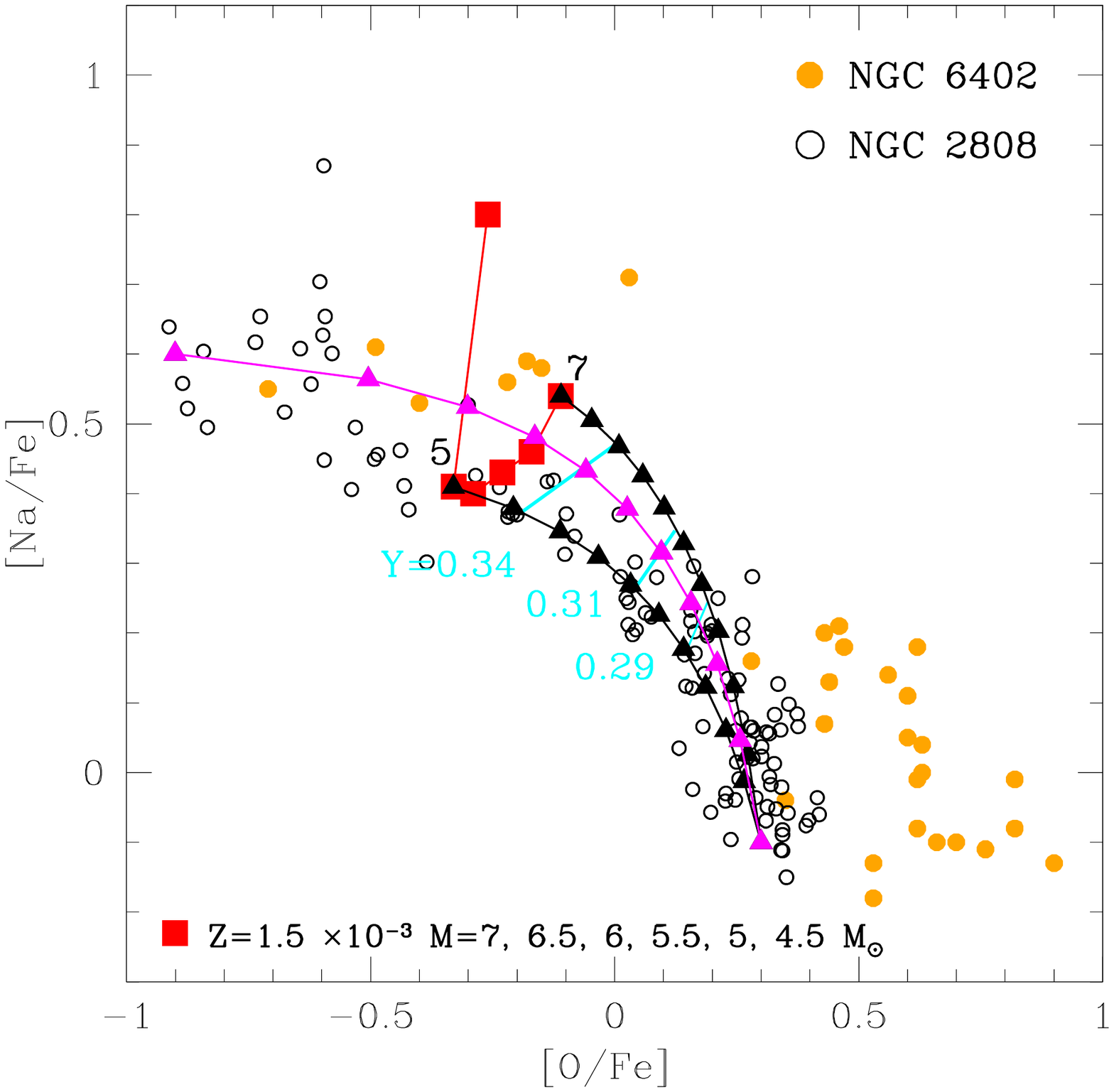}
\caption{The Na--O data for NGC\,2808 (black open circles) and for NGC\,6402 (orange dots) are shown together with the average Na--O content in the ejecta of AGBs and super--AGBs having Z=1.5$\times 10^{-3}$ shown as red squares. The top right square refers to 7\msun, and the last square on the left represents the abundances in the 4.5\msun\ ejecta. The black lines with triangles represent the abundances obtained by diluting the ejecta of the 7\msun\ and of the 5\msun\ with increasing amounts of a pristine gas having the abundance [Na/Fe]=-0.15 and [O/Fe]=0.3. Actually oxygen in the 1G of NGC\,6402 looks much larger on average, with respect to the Carretta 2015 determination of the 1G values in NGC\,2808 and in many other clusters (Carretta 2009a). The diagonal cyan lines show the helium content along the dilution lines --see text. }
\label{fig-ona}
\end{figure}
Concomitantly, the helium content of the E populations must be equal to the abundance in the pure ejecta of these stars, and all models agree that, for general structural reasons, this must be in the range Y=0.34--0.38 \citep{ventura2010}. Note that this prediction is confirmed by the indirect derivation of the helium content in the ``blue main sequence" of \ocen\ \citep[e.g.][]{tailo2016} and NGC\,2808 \citep{dc2008} and it is one of the successes of the AGB model: while the SMS model {\it has to require} that the main sequence evolution stops  at a very specific time, so  that the average helium after dilution is close to that required by observations, the AGB model predicts this limit. \\

Another straightforward prediction of the AGB model is that the nucleosynthesis, for the same degree of dilution, should be more advanced in clusters of smaller metallicity \citep{ventura2013}. Lower amounts of metals imply lower opacities in the AGB envelopes, and a larger temperature at the basis of the envelope where p-capture nucleosynthesis takes place. In particular, at higher metallicity the ON chain and p-captures on Mg isotopes are less efficient. This naturally produces the ``vertical" Na--O relations of the bulge clusters having the largest GC metals \citep[][and references therein]{munoz2020}.  \\

In this work, our attention must anyway consider the crucial issue not fully solved in the models, namely the oxygen and sodium yields. A thorough discussion is given in \cite{renzini2015}, and can be summarized by saying that, in the range of the HBB temperatures of massive AGBs, the p-captures on sodium destroy it, while at the same time p-captures on oxygen nuclei convert it to nitrogen. The interplay between these two concomitant reactions is such that sodium is depleted, if the star lives long enough to pursue a strong oxygen depletion. 
Thus, strong mass loss (as in the super--AGBs) preserves sodium, but does not destroy enough oxygen, while smaller rates of mass loss (as in the massive AGBs) can deplete more oxygen, but preserve less sodium. The resulting O-Na relation is shown by the (red) hook line in Figure\,\ref{fig-ona}. The yields shown as red squares are from the extension of the work by \cite{ventura2013} to a metal mass fraction Z=1.5$\times 10^{-3}$ ([Fe/H]$\sim$--1.3) \citep{dellagli2018} close to the metallicity of both NGC\,2808 and NGC\,6402. The upper right point red square results from the evolution of 7\msun, smaller masses follow until the 5\msun\ model, which is the mass of maximum oxygen depletion, while the 4.5\msun\ model again displays larger sodium and oxygen abundances in the ejecta. \\

The yields in the range 5--7\msun\ are close enough to the cooler points of the E group in \cite{johnson2019}. The three most oxygen depleted giants ([O/Fe]$\lesssim -0.4$) are out of the curves, as well as the E stars in NGC\,2808. In the context of the AGB model, an additional oxygen reduction is attributed \citep{dercole2012, dantona2016} to anomalous ``deep mixing" in giants \citep{dantonaventura2007}, active when a high helium content reduces the discontinuity in the hydrogen profile left by the maximum deepening of convection at the first dredge up. Notice that this interpretation {\it requires} that these stars are born with a high helium content, such as the maximum helium yield of the massive AGB and super-AGB evolution discussed above.\\

In summary, the yields are reasonable for the stars of the E group with lower O--depletion, but require an additional hypothesis to describe the three most extreme giants, according to today's state of the art. 

\subsection{A mixed model}
As discussed above, only SMS and AGB stars reach temperatures high enough to account  for the chemical composition of 2G extreme stars, even if with the mentioned uncertainties.  In addition, other dynamical events may concur to determine the present day stellar content of NGC\,6402. In particular we can consider whether the cluster is the result of the merging of different clusters \citep[e.g.][]{sugimoto-makino1989, vandenbergh1996,amaroseoane2013}. \cite{searlezinn1978} were the first to argue that some fraction of the galactic GCs has an external origin, and that the outer halo GC system is due to the merger and accretion of `protogalactic fragments'\footnote{Today this hypothesis has a part in the cosmological context of hierarchical structure formation \citep[e.g.][]{forbes2018}.} and \cite{searle1977} even argued that merging of clusters could be possible within these fragments.  Mergings occurring in the galactic thick disk are modelled today \citep[e.g.][]{khoperskov2018}; they may be relevant to understand the difference in metallicity and the difference in age by many gigayears between the two populations of the bulge clusters Terzan\,5 \citep{ferraro2016} and Liller\,1 \citep{ferraro2021}. Cluster mergers have also been explored as a possible solution to the existence of iron-complex clusters and the iron-complex clusters as well \citep[e.g.][]{gavagnin2016}. \\
In the case of NGC\,6402, there are two aspects for which it is useful to consider the merging hypothesis: 

1) if the 2G extreme stars are manifactured by the SMS ejecta  in one cluster and the 2G mild stars are born from AGB polluted gas in another one, the presence of the gap in abundances has a simple solution; this may help address the primary SMS problem that very high stellar masses are required (see Sect.\ref{subsec:SMS}); 

2) merging could explain the presence of a small fraction of  (likely) metal-richer stars identified in the ChM (group 2G$_{\rm D}$).

\section{Discussion}
Taken at face value, the high dispersion spectra observations give partially contradictory results about how extreme the E composition stars are in NGC\,6402.  Magnesium abundances are reduced at most by $\sim$0.2\,dex \citep[see Fig. 8 in ][]{johnson2019}, to be compared with the $\sim$0.5\,dex of the E stars in NGC\,2808 \citep{carretta2015}, so we should 
conclude that the E stars in NGC\,6402 are not as extreme as in NGC\,2808. \\

On the other hand, the oxygen abundances in the E sample of both clusters have formally a similar logarithmic depletion (by $\sim 1.2$\,dex) with respect to the 1G (or P1) abundances, indicating a similarly ``extreme" nuclear processing. This result, anyway, is mostly due to the large difference between the [O/Fe]$_{\rm 1G}$$\sim$0.3 of NGC\,2808 \citep{carretta2009a,carretta2015}, and the [O/Fe]$_{\rm 1G}$$\sim$0.6 of NGC\,6402 \citep{johnson2019} ---see Fig.\,\ref{fig-ona}. The zero point of the 1G is very important for the depletion models, as the ON cycle depletes oxygen in proportion to its initial content, so that, in a first approximation, stellar models having the same physical conditions (e.g. HBB temperature) but different initial oxygen abundance, have about the same logarithmic decrease in oxygen. \\

These contrasting results, together with the result of the present work, lead us to assume, for the present analysis, that there is a zero point difference between the 1G abundances by \cite{carretta2015} and by \cite{johnson2019}, and that we can discuss the depletion of oxygen in NGC\,6402 assuming that its 1G abundance were a bare [O/Fe]$\simeq$0.3. With this assumption, the E stars in NGC\,6402 are not oxygen depleted as much as those in NGC\,2808.

\subsection{Helium in the extreme population of NGC 6402}
The present work shows that the multiple populations hosted in NGC\,6402 are not so extreme as the multiple populations in the prototype cluster NGC\,2808. Three direct comparisons ---simplified by the very similar metallicity of the two clusters--- show this:
\begin{enumerate}
     \item the ChM of NGC\,6402 does not contain stars in the location of the E group in NGC\,2808 \citep{milone2015} (see Sect.\,\ref{sec:CMDs});
    \item the comparison of the CM diagrams of the two clusters shows that the blue MS of NGC\,2808, identified with a population having Y$\sim$0.35-0.38 \citep{dantona2005, piotto2007} is much bluer than the extreme MS in NGC\,6402, which can be fit with Y$\sim$0.31 (see Sect.\,\ref{sub:extremes});
    \item the HB of NGC\,6402 does not include the well populated blue hook, generally modelled as the site of post--RGB evolution of the high helium population \citep{dc2004,lee2005,dc2008, tailo2015}. 
    If the reddest HB is modelled by assuming that it is populated by 1G stars with Y=0.25, the extreme HB can be fitted with stars having Y=0.31. Consideration on the extra--mass loss needed to fit 2G HB stars \citep{tailo2020} shows that this value Y=0.31 may be an upper limit to the helium content of the hot HB stars (see Sect.\,\ref{sub:extremes});
\end{enumerate}
We first investigate whether the AGB model and/or the SMS model are in qualitative and/or quantitative agreement with this result.\\
The conclusions inferred in the context of the AGB scenario are illustrated 
 in Fig.\,\ref{fig-ona}: the three  cyan lines mark the O--Na locus where the dilution provides a helium abundance of Y=0.34, 0.31 and 0.29. It is clear that only gas diluted with 40-50\% of pristine gas has Y=0.31, and consequently its O abundance is much larger than in the E stars of NGC\,6402, and viceversa for the Na abundance. The E stars oxygen is reproduced only by undiluted models (at Y$\sim 0.35$), or by models assuming deep extra-mixing, which, anyway, also requires a similarly high initial helium \citep{dantonaventura2007}. Further, the Y=0.31 locus is in the middle of the O--Na region where there is a gap in the \cite{johnson2019} data.\\
 
In order to obtain consistency with the data, we need a much stronger O--depletion in the yields (such as those obtained for smaller metallicity) but preserving the sodium. This problem has never been solved in the models \citep{renzini2015, dantona2016} and brings us back to a key point of discrepancy in the AGB model. \\

For the SMS model, based on the results shown by \cite{denissenkov2015} for M\,13 (their Figure 1 and Table 1), the location of giants with [O/Fe] in the range from --0.5 to --0.1 is obtained by dilution with 10--20\% of pristine gas, from the pure ejecta having Y=0.384. Then, also in this case these stars should have Y=0.357--0.37. 
We can not exclude that there could be a combination of fine tuned SMS conditions (central temperature of H--burning ---or SMS mass--- helium core abundance at the stop of the core evolution and gas shedding) which allows a smaller helium content concomitant with the O--Na abundances of the E population. Ad hoc computation of nucleosynthesis such as those presented in \cite{prantzos2017} could be useful to explore the range of parameters and clarify whether the `mixed' SMS+AGB model by \cite{johnson2019} can be considered feasible from the helium point of view.

\begin{figure}
\centering
\includegraphics[width=9.cm, trim={0.7cm 6.5cm 1.5cm 4cm},clip]  {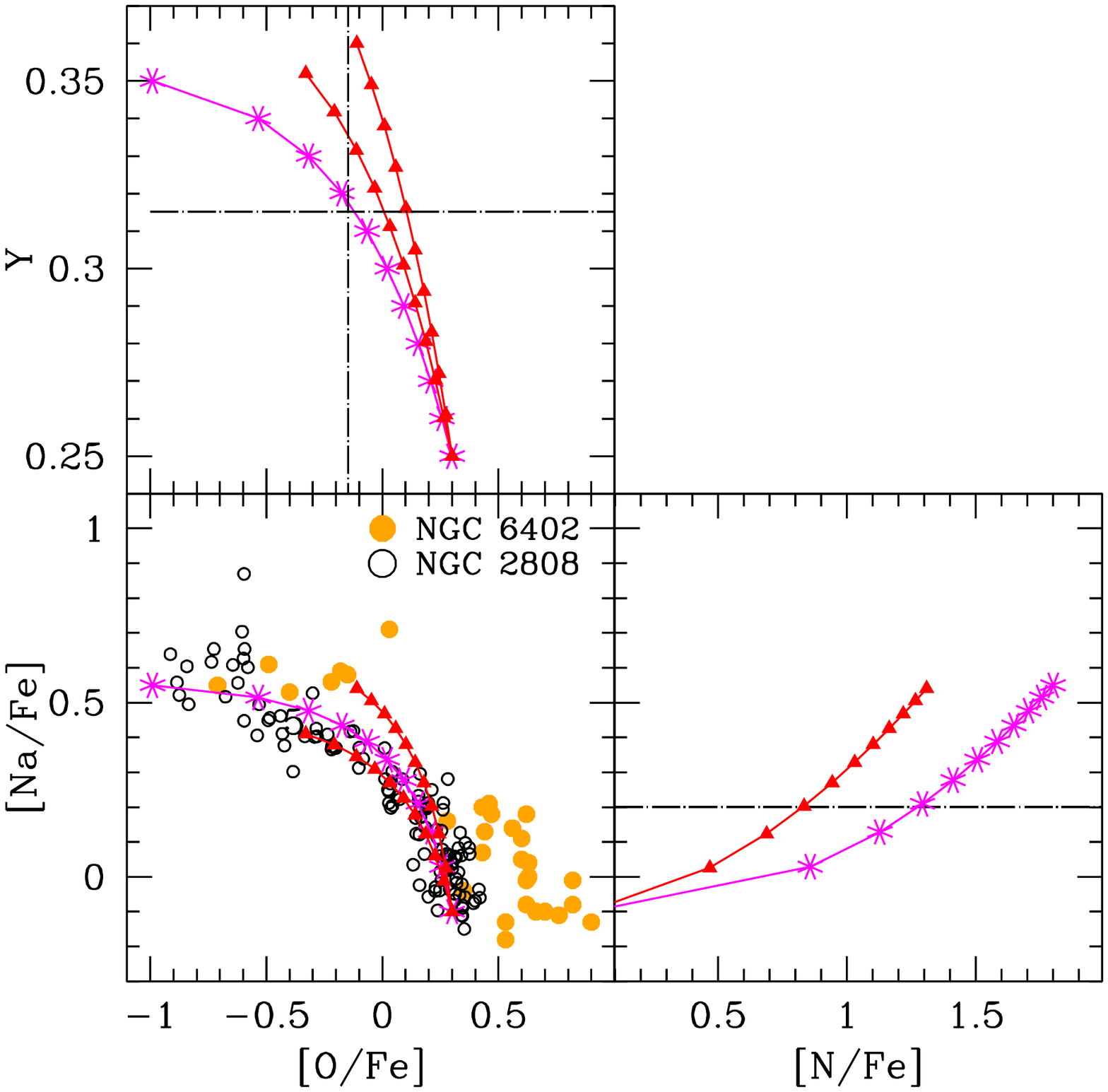}
\caption{The bottom left panel shows the Na--O data for NGC\,2808 and for NGC\,6402 together with the dilution curves of the models (see the caption of Fig.\,\ref{fig-ona}) and an ideal dilution curve (magenta line with stars) starting at [O/Fe]$=-1.0$\ and [Na/Fe]=0.55. The top panel shows the helium content along the dilution curves as function of [O/Fe] and the right bottom panel shows the [N/Fe] abundance along the dilution curve. The [N/Fe] of the ejecta having [O/Fe]=-1 in this context is roughly 1.8, 0.6\,dex larger than the [N/Fe] of the model of 5\msun\ used as comparison. 
}
\label{fig-heon}
\end{figure}

\subsection{Helium and the oxygen and nitrogen abundances: what is required from models}
Fig.\,\ref{fig-heon} shows what is needed by a successful model to conform to the results obtained from the analysis of NGC\,6402 HST data. Whatever the model (a modified AGB or  SMS  model) the ejecta of {\it at least some} of these stars must have a very low oxygen abundance (say $\sim$5\% of the pristine one), a reasonably large sodium abundance (e.g. [Na/Fe]=0.55, as chosen in the figure), but, also, a helium abundance not exceeding Y$\sim$0.35 (as found in the massive AGBs). Only in this case it is possible to have a helium abundance of $\sim$0.315 by diluting with pristine gas, and still reproduce the location in the O-Na plane of the bulk of the E stars in \cite{johnson2019} sample, as shown in the top panel. \\
New modelling is needed to understand whether the chemistry of the ejecta of super-AGB, AGB and SMS can be made compatible with this requirement.\\

The bottom right panel of Fig.\,\ref{fig-heon} shows instead the plausible run of nitrogen in the same ideal model. Indeed, if oxygen is reduced to 1/20 in the processing, more oxygen has been processed to nitrogen and the ejecta reach a value [N/Fe]$\sim 1.8$, to be compared to the value of $\sim$1.3 corresponding to a reduction of $\delta$[O/Fe]$\sim$0.4. This has an interesting consequence on our understanding of the HST results, as we show in the next section.  

\subsection{The ChM and the ``lack" of the intermediate population}

The comparison between the ChM of NGC\,6402 (black) with the ChM of NGC\,2808 (red) is shown in Fig.\,\ref{fig:Comp2808chm}, and highlights how powerful photometry can be at providing a complementary vision to the high dispersion spectroscopic results. 
The large gaps in [O/Fe], [Na/Fe], and [Al/Fe] distributions found in the spectra suggest that at most 10\% of the total population is contained in the ``intermediate" range (the I (or D) population). This is in partial agreement with the UV photometry, as the ChM shows overdensities for the E group and for the stars closer to the first population. 
In the \cite{dantona2016} scheme for the timing of formation of the different MPs, the gap in abundances is obtained if, after the formation of the moderately diluted 2G extreme E star group, re-accretion of pristine gas is fast enough that the late forming stars belong to the very diluted P2 (or 2G mild) group.
On the other hand, \citet{johnson2019} interpret the gap as a possible evidence of delayed formation for intermediate composition stars. Thus the birth of the second population would consist of two distinct phases, an early one to form the 2G extreme stars, and a very late one to form the P2 (or 2G mild) stars, and suggests the possibility that there are two different classes of pollutors. 
The merging of two clusters {\it with very close initial abundance} would provide a natural site for this occurrence.

\subsection{The group 2G$_{\rm D}$}
The ChM in Fig.\,\ref{fig:ChMs} and its comparison with NGC\,2808 in the upper left panel of Fig.\,\ref{fig:Comp2808chm}  shows clearly that the extreme 2G  in NGC\,2808 extends beyond the extreme 2G of NGC\,6402.  At the same time, it shows that NGC\,6402 includes a group of stars which are typical of Type\,II or iron--complex clusters. Lacking spectroscopic evidence, we can speculate that $\sim$9\% of the cluster stars are enriched in metals. Notice that they are 2G stars,  without a clear 1G counterpart. Also NGC\,2808  possibly hosts stars with higher metallicity (group A in Fig.\,\ref{fig:Comp2808chm}) but, on the contrary,  {\it they belong to 1G}. This must be telling us something about the pollution mechanism. In the \cite{dantona2016} model, the higher metallicity groups are born from gas contaminated by the first SN\,Ia explosions, before these explosions fully expel the gas out of the GC putting an end to 2G star formation. The timing of this event helps to explain why the iron-richer groups are often s-process rich, being polluted by the ejecta of AGBs of smaller mass, where s-process enhancements are due to the prolonged third dredge up phases. In the case of NGC\,6402,  the SN\,Ia pollution must have occurred mainly on p-processed gas, while pollution took place within a pristine gas region in NGC\,2808.\\
As the  2G$_{\rm D}$\ group is mainly 2G,  it is unlikely that it is a signature of merging. 
Further  speculation  is  beyond the scope  of  this  work.   The  results  presented here  for NGC 6402 call for additional efforts in the development of new theoretical models.

\subsection{The lack of a gap in the ChM}
We remark here that the lack of a ``gap" in the ChM map is not in contradiction with a possible gap in some of the light elements abundances, because the ordinate in the ChM mostly traces {\it the nitrogen abundance}, which can be more continuous than Na or Al, {\it if the p-processed gas comes from progenitors having different nucleosynthesis patterns}. From Fig.\,\ref{fig-heon}, if there are indeed yields as low in oxygen to justify the low value of Y=0.315 for the E population, necessarily their nitrogen abundance in the ejecta is much higher, up to a factor 2.5 (or 0.4 in the log) larger than the nitrogen in ejecta which deplete less  oxygen, such as in the 5--7\msun AGBs models. The right bottom panel of Fig.\,\ref{fig-heon} shows then that stars with similar sodium abundance (e.g.[Na/Fe]=0.2) may differ by up to 0.5\,dex in nitrogen. Thus evolution of stars with intermediate oxygen depletion (and different nitrogen) may result in a continuous coverage of the ChM.  Spectroscopy of stars in the less populated region of the ChM would be very useful, because they can help to discriminate between stars formed by ejecta with different compositions.

\section{Conclusions}
We have presented multi-band photometry of the massive GC NGC\,6402 based on multi-band {\it HST} observations collected as part of GO-16283 (PI.\,F.\,D'Antona). Our photometric diagrams have been combined with results from high-resolution spectroscopy \citep[][]{johnson2019} to constrain the formation scenarios of multiple populations in GCs. \\
 
 The huge gap in the distribution of some light elements (O, Mg, Al and Na) found by \cite{johnson2019} is certainly one of the most-intriguing properties of multiple populations in NGC\,6402. One of the main objectives of our work consists in exploring whether NGC\,6402 lacks stellar populations with intermediate chemical composition, based on a large stellar sample.
 To do this, we exploit the ChM, which is the photometric diagram that has been previously used to identify and characterize multiple populations in 57 Galactic GCs as part of the {\it HST} UV Legacy Survey of GCs \citep[][]{piotto2015, milone2015, milone2017chromo}.
  The main results of the new observations concerning the ChM are the following:

\begin{enumerate}
    \item the ChM do not show stars with extreme chemical compositions as observed in some massive GCs (e.g. NGC\,2808 and $\omega$\,Cen).
    Notably, NGC\,6402 lacks the E group stars of the prototype cluster NGC\,2808;
    \item the ChM contains a group of 2G stars (dubbed 2G$_{\rm D}$) on the red side of the diagram, indicating the presence of a small group of stars ($\sim 8$\%) having higher metallicity: NGC\,6402 is then a ``Type II" cluster, in the definition by \cite{milone2017chromo};
    \item the ChM does not show any gaps. There is a region less populated in the middle region, between the extreme groups (2G$_{\rm C}$\ and 2G$_{\rm A}$).
    When we locate on the ChM a few giants in common with the spectroscopic sample, we find that there is an extreme star also among the less extreme group 2G$_{\rm B}$.
\end{enumerate}
Thus it is very important to broaden the spectroscopic investigation of this cluster and study targets in the middle of the ChM.

An independent result of the observation was obtained in the parallel field data: the color magnitude diagram shows indeed a split MS, confirming that we are in the presence of a dichotomy in the helium abundance of the cluster stars. Therefore, the spectroscopic gap is fully confirmed by the presence of the MS gap. These results indicate that the cluster has undergone two stages of 2G star formation, the first one from matter strongly contaminated with p-processed elements and significantly rich in helium, the second one from matter ---not necessarily sharing the same heavy p-processing of the first event--- heavily diluted with pristine gas, so that  both the helium content and the abundances of light elements remain either standard or close to standard. \\

Already a simple superposition of the parallel field data of NGC\,2808 and  NGC\,6402  shows that the blue MS of the latter cluster is much less extreme, and thus its helium content is much smaller than the Y$\sim$0.35 attributed to the prototype cluster. In fact, the analysis of the MS data provides Y$\sim$0.31 for this blue MS (and notice that also the ChM analysis provides a similarly ``low" Y=030 for the extreme stars). This helium content for the extreme population is also consistent with the analysis of the HB. \\

The different samples examined give fractions of stars in the three main groups defined by spectroscopy (1G, mild 2G and extreme 2G) reasonably compatible with each other, apart from the HB stars, which has far too few ``red" stars, which should correspond to the 1G plus mild 2G sample. We will address this problem in a future study. \\

We find that the determination of a helium content Y$\sim 0.31$\ for the 2G extreme stars is extremely useful, because it confirms the composition gap, and allows a more demanding comparison between models and abundances. We examine both the SMS model and the AGB model, and conclude that the present O--Na abundance patterns of the AGB yield for the metallicity of this cluster (Z=1.5$\times 10^{-3}$) are not compatible with Y$\sim 0.31$. Also SMS present models seem not compatible with the data, but we leave some space to further exploration of the parameters space ({\it i:}SMS mass ---or core temperature of nuclear processing--- and {\it ii:} helium content at which the evolution of the SMS is artificially stopped). \\

Concerning the AGB model, we conclude that models allowing for a larger oxygen depletion preserving sodium are needed to be compatible with the data. Whether and under which hypotheses this is possible is to be explored. Anyway, it is difficult to maintain the hypothesis that the most extreme oxygen abundances found in GCs are simply due to deep mixing in the high helium red giants, favored by the small chemical discontinuity left by convection at the first dredge up \citep{dantonaventura2007}. \\

One bonus of having established the degree of oxygen depletion necessary for the 1G ejecta polluting the extreme star sample is the following:  the pollutors of the 2G population will probably cover a range of oxygen depletion, from the most extreme value [O/Fe]$\sim -1$, to the moderate depletion shown by the AGB models. Consequently, after dilution with pristine gas, the gas forming the 2G mild stars  may also {\it cover a range of nitrogen abundances} able to smooth down any gap in the ChM.

\section*{Acknowledgements}
This work has received funding from the European Research Council (ERC) under the European Union's Horizon 2020 research innovation programme (Grant Agreement ERC-StG 2016, No 716082 'GALFOR', PI: Milone, http://progetti.dfa.unipd.it/GALFOR).
APM acknowledges support from MIUR through the FARE project R164RM93XW SEMPLICE (PI: Milone). APM has been supported by MIUR under PRIN program 2017Z2HSMF (PI: Bedin). MT acknowledges support from the European Research Council Consolidator Grant funding scheme (project ASTEROCHRONOMETRY, G.A. n. 72293, http://www.asterochronometry.eu).
We thank the referee for suggestions and for a very careful reading of the manuscript.


\bibliographystyle{aasjournal}
\bibliography{game2018} 


\end{document}